\documentclass{jpp}
\usepackage{graphicx}
\usepackage{epstopdf, epsfig}

\usepackage{todonotes}

\renewcommand{\deg}{^{\circ}}
\renewcommand{\th}{\vartheta}
\newcommand{\vcr}{v_{\rm iso}}
\newcommand{\ncr}{n_{\rm CR}}
\renewcommand{\vcr}{v_{\rm CR}}
\newcommand{\jcr}{J_{\rm CR}}
\newcommand{\esh}{E_{\rm sh}}
\newcommand{\vsh}{v_{\rm sh}}

\newcommand{\epsp}{\varepsilon_{\rm p}}
\newcommand{\epscr}{\varepsilon_{\rm CR}}

\shorttitle{DSRA}
\shortauthor{D. Caprioli, H. Zhang, A. Spitkovsky}

\title{Diffusive Shock Re-Acceleration}

\author{Damiano Caprioli\aff{1,2}
  \corresp{\email{caprioli@uchicago.edu}},
  Horace Zhang\aff{2}
 \and Anatoly Spitkovsky\aff{2}}

\affiliation{\aff{1}Department of Astronomy and Astrophysics, University of Chicago,\\
5640 S Ellis Ave, Chicago, IL 60637, United States
\aff{2}Department of Astrophysical Sciences,\\ Princeton University, 4 Ivy Ln., Princeton, NJ 08544, United States}

\begin{document}

\maketitle

\tableofcontents

\newpage
\begin{abstract}
We have performed 2D hybrid simulations of non-relativistic collisionless shocks in the presence of pre-existing energetic particles (``seeds'');
such a study applies, for instance, to the re-acceleration of Galactic cosmic rays (CRs) in supernova remnant (SNR) shocks and solar wind energetic particles in heliospheric shocks.
Energetic particles can be effectively reflected and accelerated regardless of shock inclination via a process that we call \emph{diffusive shock re-acceleration}.
We find that re-accelerated seeds can drive the streaming instability in the shock upstream and produce effective magnetic field amplification. 
This can eventually trigger the injection of thermal protons even at oblique shocks that ordinarily cannot inject thermal particles.
We characterize the current in reflected seeds, finding that it tends to a universal value $J\simeq e\ncr\vsh$, where $e\ncr$ is the seed charge density and $\vsh$ is the shock velocity.
When applying our results to SNRs, we find that the re-acceleration of Galactic CRs can excite the Bell instability to non-linear levels in less than $\sim 10$ yr, thereby providing a minimum level of magnetic field amplification for any SNR shock.
Finally, we discuss the relevance of diffusive shock re-acceleration also for other environments, such as heliospheric shocks, Galactic superbubbles, and clusters of galaxies.
\end{abstract}

\section{Introduction}
Collisionless shocks are ubiquitous in space and astrophysical environments and are often associated with non-thermal particle acceleration and emission.
Important examples are non-relativistic SNR shocks, which are widely regarded as sources of Galactic CRs \citep{tycho,crspectrum}, and heliospheric shocks, where  particle acceleration can be investigated via in-situ spacecraft observations.

In the past few years, modern supercomputers have opened a new window for investigating the non-linear interaction between accelerated particles and electromagnetic fluctuations from first principles via kinetic particle-in-cell simulations. 
A crucial contribution to the present understanding of particle acceleration at non-relativistic shocks came from hybrid (kinetic ions--fluid electrons) simulations, which allow to fully capture the ion dynamics and the development of plasma instabilities at a fraction of computational cost required to also follow the electron dynamics with full particle-in-cells simulations.

In a recent series of papers, we used hybrid simulations to perform a comprehensive analysis of the ion acceleration at collisionless shocks as a function of the strength and topology of the pre-shock magnetic field \citep{DSA}; 
the nature of ion-driven instabilities \citep{MFA}; 
the transport of energetic ions in self-generated magnetic turbulence \citep{diffusion}; 
the injection of thermal ions into the acceleration process \citep{injection}; 
and the acceleration of ions with arbitrary mass-to-charge ratio \citep{AZ}.
For an outline of some phenomenological implication of these results, see the recent review by \cite{icrc15}.

In this paper we want to generalize such results to situations in which the shock runs into a medium that is already filled with energetic seed particles, as is typically the case in the interstellar medium or the solar wind. 
This is a crucial step towards a better understanding of interstellar and heliospheric shocks, whose observed phenomenology is not always explained in terms of the most common acceleration mechanism, namely diffusive shock acceleration (DSA) \citep[][]{blandford-ostriker78,bell78a}.
After a brief introduction about how DSA occurs for shocks propagating at different angles with respect to the large-scale magnetic field, in \S\ref{sec:seeds} we discuss the injection and acceleration of pre-existing energetic seeds for oblique shocks.
In particular, we address the triggering of the Bell instability driven by the current in reflected CRs, which  provides crucial back-reaction on the global shock structure.
In \S\ref{sec:eps} we study the (re)acceleration efficiency of both seeds and thermal protons for different shock inclinations, comparing the results with and without energetic seeds.
The general properties of the current of reflected CRs are worked out and discussed in \S\ref{sec:jcr}.
Finally, in \S\ref{sec:SNR} were-acc put our findings in the context of re-acceleration of Galactic CRs seeds in SNR shocks before concluding in \S\ref{sec:conc}.

\section{Hybrid Simulations with Energetic Seeds\label{sec:seeds}}

\subsection{DSA and Shock Inclination}
A crucial parameter that controls how efficiently a shock can channel kinetic energy into non-thermal particles is its inclination, defined by the angle $\th$ between the direction of the large-scale magnetic field ${\bf B}_0$ and the shock normal, such that $\th=0\deg$ ($\th=90\deg$) corresponds to parallel (perpendicular) shocks; 
in the following we also use \emph{oblique} for shocks with $45\deg\lesssim\th\lesssim 70\deg$.
${\bf B}_0$ is chosen to be in the simulation plane because such a configuration returns results consistent with 3D setups \citep[e.g.,][]{ss11,DSA}.

In \cite{DSA} we found that the acceleration of thermal ions is efficient at quasi-parallel shocks: more than 10\% of the shock ram kinetic energy can be converted into energetic particles with the universal power-law tail predicted by the DSA theory. 
For oblique shocks, the acceleration efficiency is reduced and becomes negligible above $\th=60\deg$.
The reason for such a behavior is discussed in \cite{injection}, where we studied how ion injection is controlled by reflection off the shock electrostatic barrier, which oscillates on a cyclotron timescale, and the shock inclination. 
In order for protons to be injected into DSA, they need to achieve a minimum energy $E_{\rm inj}(\th)$, and $E_{\rm inj}(\th)$ is an increasing function of $\th$.
In order to achieve the injection energy, protons must be reflected by the shock (and gain energy via shock drift acceleration, SDA) a certain number of times, but at each encounter with the reforming shock barrier they have a probability of $\sim 75\%$ to be advected away downstream;
therefore, the fraction of particles that can undergo $N$ SDA cycles is typically $\sim 0.25^N$.
We also found that for $\th\lesssim45\deg$ the injection energy is $\sim 10\esh$, which is achieved with $N\simeq3$ SDA cycles, returning an injection fraction of $\sim 1\%$, in good agreement with simulations.
Larger shock inclinations, however, require $N\gtrsim 3$ to reach $E_{\rm inj}$ and the fraction of injected ions drops exponentially with $\th$.
When the injection fraction is about 1\% by number, the current in energetic particles is large enough to drive a very effective amplification of the initial magnetic field;
for $\th\gtrsim 50\deg$, instead, the fraction of injected thermal particles is much smaller and the current is too weak to amplify the field \citep{MFA}.
The net result is that quasi-parallel shocks can \emph{spontaneously} inject particles from thermal energies, which leads to a very efficient DSA and magnetic field amplification, while more oblique shocks cannot.    

\subsection{Hybrid Simulation Setup}
In this section we investigate how the presence of seeds with an initial energy exceeding $E_{\rm inj}$ may overcome the injection problem for oblique shocks.
We use the code \emph{dHybrid} \citep{gargate+07} to run simulations of non-relativistic shocks including pre-energized particles. 
Simulations are 2D, but we account for the three spatial components of the particle momentum and of the electric and magnetic fields. 
As usual, we normalize lengths to the proton skin depth, $c/\omega_p$, where $c$ is the speed of light and $\omega_p\equiv \sqrt{4\pi n_p e^2/m}$ is the proton plasma frequency, with $m$, $e$ and $n_p$ the proton mass, charge and number density.
Time is measured in units of inverse proton cyclotron frequency, $\omega_c^{-1}\equiv mc/eB_0$, where $B_0$ is the strength of the initial magnetic field.
Finally, velocities are normalized to the Alfv\'en speed $v_A\equiv B/\sqrt{4\pi m n}$, and energies are given in units of $\esh\equiv m\vsh^2/2$, with $\vsh$ the velocity of the upstream fluid in the downstream frame, which is also the simulation frame.
Shocks are produced by sending a supersonic flow of velocity $\vsh$ against a reflecting wall and are characterized by their sonic and Alfv\'enic Mach numbers, $M_s\equiv\vsh/c_s$, $M_A\equiv\vsh/v_A$, with $c_s$ the sound speed\footnote{Note that $\vsh$ defines the upstream flow velocity in the downstream frame; the shock velocity in the upstream frame is slightly larger, $\vsh (1+1/r)$, where $r=\rho_d/\rho_u\simeq 4$ is the shock compression ratio, i.e., the ratio of the downstream to upstream density.}.

Finally, fluid electrons are initialized with the same temperature as ions, and have a polytropic equation of state with an effective index $\gamma_{\rm eff}(M_s)$ chosen in order to satisfy the Rankine--Hugoniot conditions for a shock where thermal equilibration between  protons and electrons is maintained across the shock (see Appendix A for details).
Note that very rapid electron--ion equilibration in the shock transition is typically seen in full PIC simulations of non-relativistic shocks  when proton-driven turbulence is present \citep[e.g.][]{electrons}.
In any case, we have checked that the main results in this work do not depend on the prescription for the electron equation of state. 

The novelty in the simulations presented here is the presence of an additional population of energetic seeds, initialized in the upstream reference frame as isotropic and with a flat distribution in each momentum component $p_i$ in the range $-m \vcr\leq p_i\leq m \vcr$, which corresponds to an average energy of $(\vcr/\vsh)^2\esh$.
In the simulation frame such a population is also drifting along with the thermal ions with velocity $\vsh$.
We refer to this component either as ``{seeds}'' or ``CRs'' throughout the paper. 
For the CRs, the left side of the box is open (not a reflective wall) to prevent the formation of an additional shock on the CR scales. 

As a benchmark run, we consider a strong shock with $M_s\simeq M_A\equiv M=30$ and $\th=60\deg$, a configuration where thermal ions are hardly injected \citep{DSA}.
The time step is chosen as $\Delta t=0.0015 \omega_c^{-1}$ and the computational box measures $[L_x,L_y]=[10^5, 500] c/\omega_p$, with two cells per ion skin depth and four particles per cell for both protons and CRs.
The CRs drift with the incoming flow into the shock and have $\vcr = 50 v_A$ and $\ncr = 0.01$, so that the energy density in CRs is negligible ($\lesssim 3\%$) with respect to the proton kinetic energy.
We will discuss how results depend on the choice of $M$, $\th$, $\vcr$, and $\ncr$ later in the paper.

\subsection{Cosmic Ray Injection and Re-Acceleration} 
\begin{figure}{}
\centering
\includegraphics[trim=5px 50px 0px 310px, clip, width= 1.0 \textwidth]{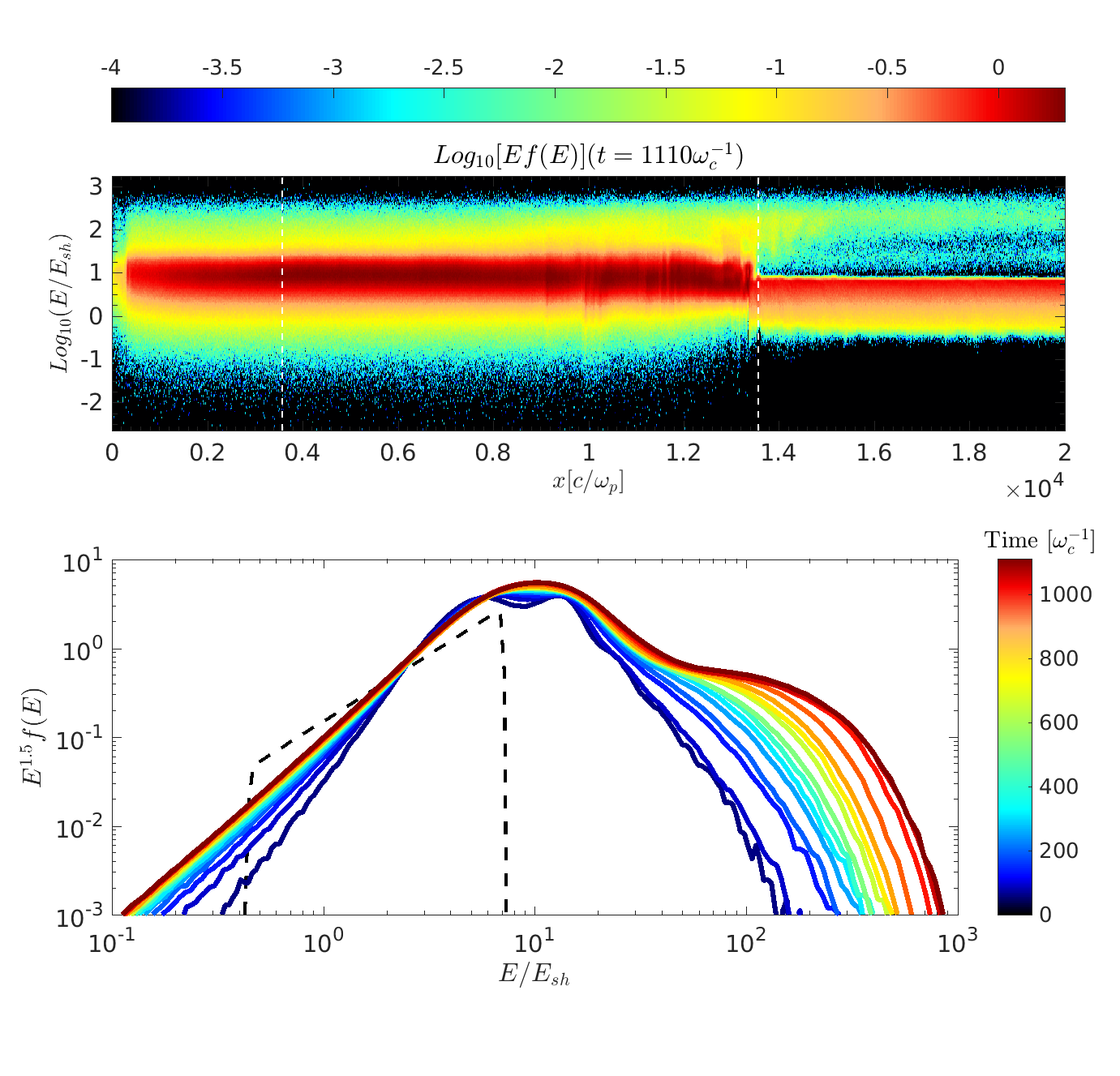}
\caption{Time evolution of the downstream CR energy spectra (see colorbar) for our benchmark shock with $\th = 60\deg$, $M = 30$, and seeds with $\vcr = 50v_A$ and $\ncr = 0.01$. 
The dashed black line shows the initial energy spectrum of CR seeds.
Spectra are multiplied by $E^{1.5}$ to demonstrate agreement with DSA theory. 
The growth of the maximum energy and the flattening of the power law tail shows that energetic CRs are injected into DSA, even in an oblique shock where thermal ions are not spontaneously injected.} \label{fig:spec}
\end{figure}

Figure \ref{fig:spec} shows that the post-shock CR spectrum (integrated over the whole downstream), initially peaked around $10\esh$, develops a DSA power-law tail whose extent (the exponential cutoff at high energies) increases with time. 
For strong shocks, the universal DSA momentum spectrum is $f(p)\propto p^{-4}$, which translates into an energy spectrum $f(E) = 4 \pi p^2 f(p) \frac{dp}{dE}$.
Since for non relativistic particles $p \propto E^{1/2}$, the universal DSA energy spectrum is $f(E) \propto E^{-1.5}$. 
The CR energy distribution $f(E)$ in figure \ref{fig:spec} consistently converges to the theoretical slope, since particles are not allowed to become relativistic in \emph{dHybrid}. 
The number fraction of CRs in the non-thermal tail is $\gtrsim10\%$, much larger than the typical fraction of $\lesssim 1\%$ of protons that get injected and accelerated via DSA.
Finally, at late times the low-energy part of the spectrum relaxes towards a Maxwellian-like distribution because of collisionless interactions mediated by the self-generated magnetic turbulence.

For such an oblique shock we do not expect an effective injection of \emph{thermal} protons into DSA, because the fraction of them that can achieve the injection energy $E_{\rm inj}$ via SDA is very small.
More precisely, particle injection into DSA requires a minimum velocity along and transverse to the shock normal  to allow particles to overrun the shock and escape upstream \citep[see][figure 4]{injection}.
CR seeds differ from thermal ions in three aspects:
\begin{itemize}
\item The shock barrier is regulated by thermal protons and cannot prevent energetic CRs from propagating between the two sides of the shock, similarly to what happens for ions with large mass/charge ratio \citep{AZ};
\item Without interaction with the shock surface, SDA does not occur and CRs can be directly injected into DSA if their velocity exceeds the one required for overrunning the shock \citep{injection};
\item Seed CRs are significantly ``hotter'' than protons, in the sense that their phase space distribution is much more isotropic than that of the supersonic thermal particles; 
therefore, CRs can impinge on the shock with larger velocities transverse to the shock normal, which enhances their chances of being reflected and overruning the shock compared to cold incoming protons. 
\end{itemize}

In our benchmark case, the combination of ${\bf \vcr}$ and ${\bf \vsh}$ gives rise to CRs impinging on the shock with energies as large as $\sim(\vcr+\vsh)^2/\vsh^2\esh\sim 7\esh$ (see the dashed line in figure \ref{fig:spec}).
After reflection at the shock, this energy increases by another factor of $\sim2$ thanks to the typical energy gain for Fermi acceleration\footnote{More precisely, for such supra-thermal particles we observe SDA.}, $\Delta E/E\approx \frac 43\vsh/v$, as illustrated by the second peak at $E\gtrsim 10\esh$ visible at early times in figure \ref{fig:spec}.
Quite intriguingly, at late times the non-power-law part of the CR distribution resembles a Maxwellian, with an effective ``temperature'' of $T_{\rm CR,eff}\simeq 15\esh$, corresponding to the characteristic energy of CRs that underwent one cycle of Fermi acceleration at their first shock encounter. 
Such a phenomenon is similar to what happens to heavy ions, which thermalize to a temperature proportional to their mass \citep{AZ}.

Quasi-isotropic particles with energy $E\gtrsim 10\esh$ can overrun the shock and be injected into DSA even for oblique shocks. 
We dub this process \emph{Diffusive Shock Re-Acceleration} (DSRA) because it slightly differs from DSA of thermal particles in terms of injection, efficiency, and spectra produced.
First, seeds are not injected via specular reflection at the shock, but rather ``leak'' back from downstream.
The fraction of injected particles is not a function of the shock strength and inclination only \citep[as it is the case in the absence of seeds and pre-existing turbulence, see][]{injection}, but it also depends on the initial velocity of the seeds.
Second, the acceleration efficiency is not an intrinsic property of the shock, but rather it is  regulated by the amount of seeds available; this means that the level of self-generated magnetic turbulence is not universal, either, which may in turn limit the maximum energy achievable.
Finally, the standard DSA prediction that the spectral index depends on the compression ratio only is violated at shocks with $\th\gtrsim 70\deg$  (\S\ref{sec:qperp}); 
shock acceleration in this regime can only occur thanks to the presence of seeds because the injection of thermal particles at these obliquities is strongly suppressed.

When seeds have an initial distribution in momentum, the resulting spectrum is expected to be the flatter between the DSA spectrum and the initial seed spectrum \citep[e.g.,][]{bell78b,blasi04}.
With mono-energetic seeds we cannot validate such a prediction directly, but we argue that it should hold because DSRA spectra are either consistent with, or steeper than, the standard DSA ones.
Above the maximum seed momentum, the shock slope should always be the one produced by DSRA.
 
\begin{figure}
\includegraphics[width= 1\textwidth]{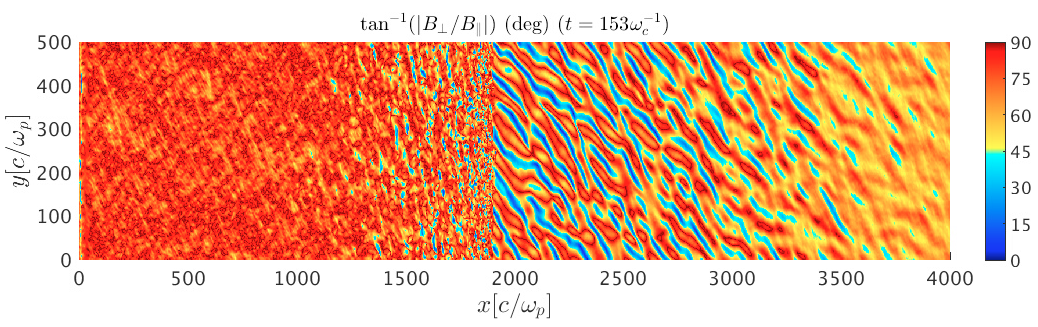}
\includegraphics[width= 1\textwidth]{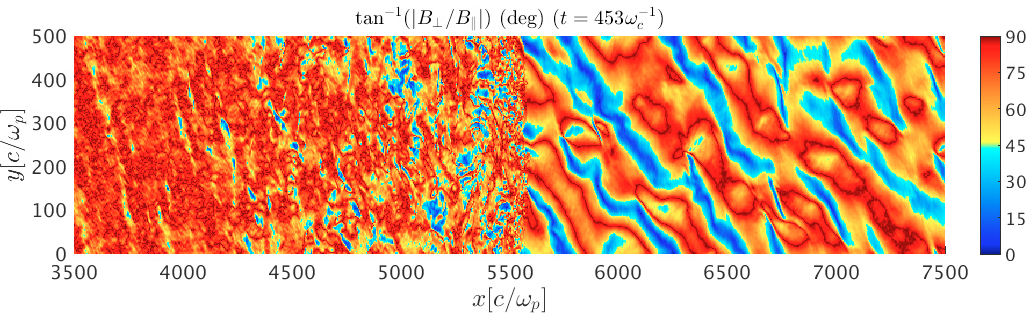}
\caption{Local magnetic field inclination around the benchmark shock with $\th = 60\deg$, $M = 30$, $\vcr = 50v_A$, and $\ncr = 0.01$ at $t\simeq 150\omega_c^{-1}$ and $t\simeq 450\omega_c^{-1}$ (top and bottom panels); 
 the upstream fluid is at $x\gtrsim 1900 c/\omega_p$ and   $x\gtrsim 5500 c/\omega_p$, respectively.
The Bell instability driven by re-accelerated CRs distorts the initial oblique field and creates quasi-parallel pockets (blue regions) where protons can be injected into DSA. 
The transverse size of such quasi-parallel filaments grows with time \citep[e.g.,][]{rb13,filam}.} \label{fig:theta}
\end{figure}

\subsection{DSRA Back-reaction\label{sec:feedback}}
DSA is always associated with an anisotropic population of energetic particles in the upstream, which drives a rearrangement or even an amplification of the background magnetic field \citep{MFA,ab09}.
The main instabilities responsible for such amplification are the resonant streaming instability \citep[e.g.,][]{skilling75a,bell78a} and the non-resonant hybrid (or Bell) instability \citep{bell04}.
The former is typical of moderately-strong shocks with $M\lesssim 30$ and saturates at quasi-linear levels of field amplification $\delta B/B_0\lesssim 1$, while for stronger shocks the Bell instability can generate very non-linear fluctuations $\delta B/B_0\gg 1$ \citep{MFA}.

We now consider the effects of the current carried by the CRs re-accelerated by the shock.
Such a strong current drives the Bell instability, which amplifies the initial $\mathbf{B}_0$ field and distorts its initially-oblique configuration, creating ``pockets'' of quasi-parallel field regions upstream of the shock.
Figure \ref{fig:theta} shows the local magnetic field inclination around the shock for $t = 153 \omega_c^{-1}$ and $t = 453 \omega_c^{-1}$ for our benchmark run. 
The initial field inclination ($\th = 60\deg$) is drastically rearranged, with quasi-parallel regions (in blue) appearing upstream of the shock in filamentary structures. 
Such structures, which start as small-wavelength perturbations, grow with time in such a way that their transverse scale is comparable with the gyroradius of the highest-energy diffusing particles, which carry most of the current \citep{filam, MFA,rb13}.
The two panels in figure \ref{fig:theta} attest to this increase in wavelength with time. 

\begin{figure}{}
\centering
\includegraphics[trim=0px 30px 0px 260px, clip, width= 1.0 \textwidth]{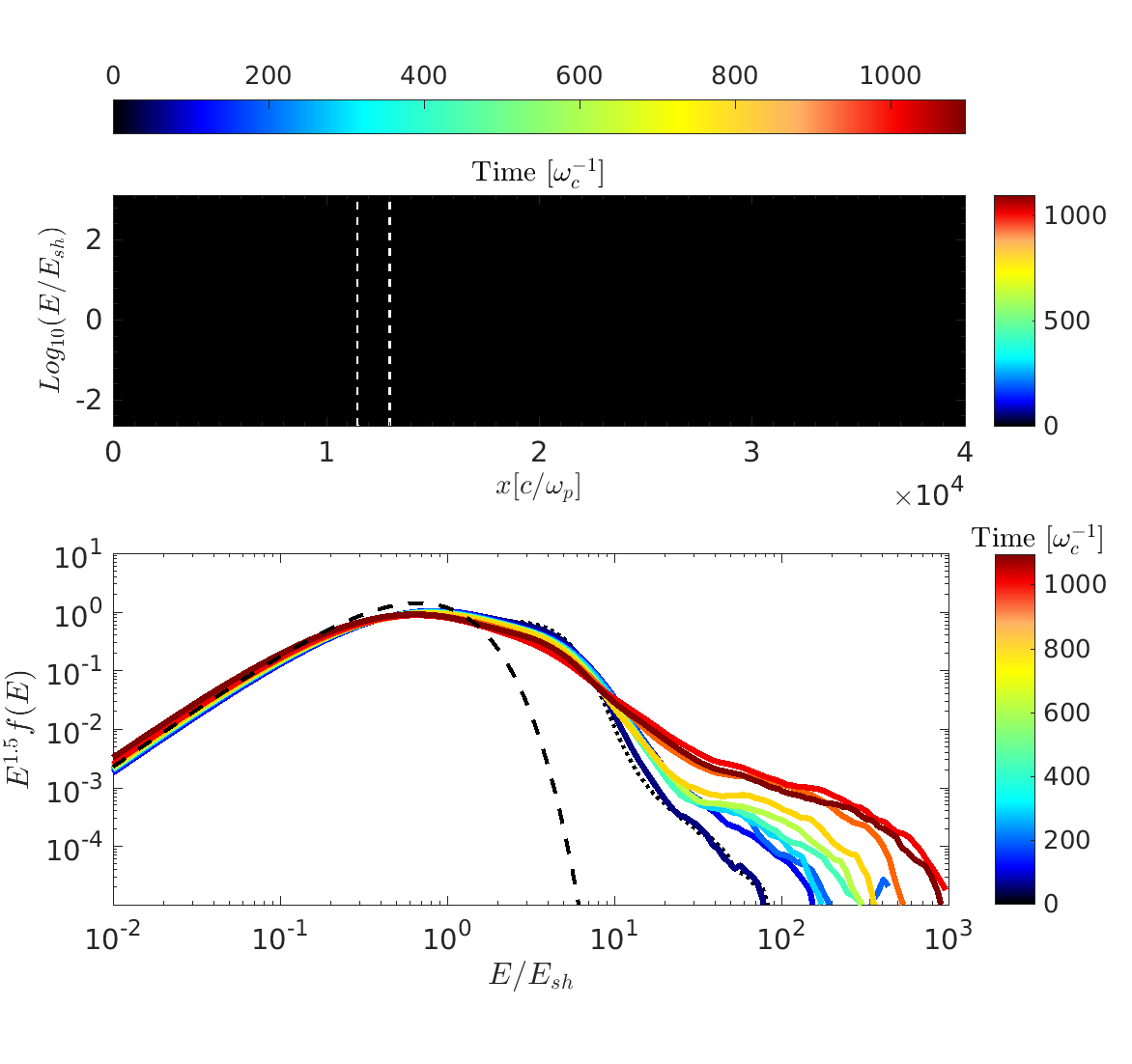}
\caption{As in figure \ref{fig:spec}, but for the initially-thermal protons. 
Protons develop a non-thermal tail after the onset of the Bell instability ($t\gtrsim 100 \omega_c^{-1}$), which opens up quasi-parallel patches at the shock surface (see figure \ref{fig:theta}) where thermal particles can be injected. 
The dashed line corresponds to the Maxwellian distribution estimated with standard Rankine--Hugoniot conditions. 
The dotted line corresponds to the proton spectrum at $t= 1000 \omega_c^{-1}$ for a shock with the same parameters, but without seeds; 
such a spectrum is virtually  indistinguishable from the one in the seeded case before the onset of the Bell instability.
Note how the supra-thermal ``bump'' (protons with energies $2\esh \lesssim E \lesssim  10\esh$) decreases with time while the non-thermal tail grows, which indicates the injection of SDA protons into DSA.} \label{fig:pspec}
\end{figure}

The presence of patches of quasi-parallel magnetic field in the non-linear stage of the Bell instability locally creates the conditions for the injection and acceleration also of thermal protons. 
Figure \ref{fig:pspec} shows the evolution of the downstream proton spectra for our benchmark run. 
The expected Maxwellian distribution (black dashed line) fits the low-energy thermal part of the spectrum well at all times. 
Supra-thermal protons with $2\esh \lesssim E \lesssim 10\esh$ are generated at the shock via SDA, as discussed in \cite{injection}, and at early times form a ``bump'', which remains stationary in the absence of CR seeds \citep{DSA}.
 With CR seeds, instead, the fraction of supra-thermal ions decreases with time, while  non-thermal power-law tail develops and tends to the expected spectrum $\propto p^{-4}$, which is achieved at $t\approx 800 \omega_c^{-1}$.
At later times the spectrum seems to be a bit steeper, which is likely due to the fact that the maximum proton energy, $E_{\rm max}$, has increased more than linearly, about a factor of 3 from $t\approx 800 \omega_c^{-1}$ to $t\approx 1000 \omega_c^{-1}$ (yellow and brown curves in figure \ref{fig:pspec}). 
In this respect, it is worth stressing that the seed maximum energy increases linearly with time since the beginning (figure \ref{fig:spec}), a  signature typical of Bohm diffusion  \citep[e.g.][]{diffusion};
conversely, protons have a smaller $E_{\rm max}$  initially, but then they can exploit the seed-generated magnetic turbulence to quickly catch up with CRs. 
At $t\approx 1000 \omega_c^{-1}$ both spectra are exponentially cut-off at $E_{\rm max}\simeq  300 \esh$.
This suggests that supra-thermal ions can be injected into the DSA process in regions where shock inclination drops below $\th\sim50\deg$. 
The fraction of injected protons, however,  is $\sim10^{-3}$, about one order of magnitude smaller than for quasi-parallel shocks.

It is useful to introduce the acceleration efficiencies $\epsp$ and $\epscr$, respectively defined as the energy density in (initially-thermal) protons with $E\gtrsim 10\esh$ and in re-accelerated seeds, normalized to the upstream bulk energy density, $n_p \vsh^2/2$, and measured behind the shock in the downstream (simulation) reference frame. 
For our benchmark run, we find $\epsp\sim 3\%$, compared to  $\lesssim 0.5\%$ for a shock with $M=30$ and $\th=60\deg$ without CR seeds \citep{DSA}. 
 
\subsection{The Onset of the Bell instability} 
To better outline the effect of CR-induced streaming instability on proton acceleration, we vary the initial CR density and check how the onset of the Bell instability and the trigger of proton injection depends on $\ncr$. 
The typical growth time of the Bell instability (in the MHD limit) is given by the reciprocal of its maximum growth rate \citep{bell04}:
\begin{equation}
\frac{1}{\Gamma_{\rm Bell}} = 2 \frac{e n_p v_A}{\jcr} \omega_c^{-1}
\end{equation}
where we introduced 
\begin{equation}
\jcr \equiv \chi e\ncr\vsh. 
\end{equation}
as the current in reflected CRs.
$\chi$ parametrizes $\jcr$ in units of $\ncr\vsh$ and represents a measure of the initial reflectivity of the shock; 
we expect it to depend on $\vcr$ and on $\th$, but not on $\ncr$, and to change in time only once non-linear effects cannot be neglected any longer.
By measuring the current in reflected CRs from simulation we find that $\chi\lesssim 1$ (see \S\ref{sec:jcr}), and in general we expect that not more than $\Xi\approx 5-10$ e-folds are needed for $\delta B/B_0$ to reach its maximum value.
Eventually, we can conclude that magnetic field amplification should reach saturation after the characteristic timescale
\begin{equation}\label{eq:tbell}
\tau_{\rm Bell} \simeq 
\frac{2\Xi}{\chi M}\frac{n_p}{\ncr}\omega_c^{-1}\simeq 
6.7~\frac{\Xi}{\chi} \left(\frac{M}{30}\right)^{-1}
\left(\frac{\ncr}{0.01}\right)^{-1} \omega_c^{-1}.
\end{equation}
For our benchmark run, $\tau_{\rm Bell}\lesssim 50\omega_c^{-1}$ after the establishment of the current in reflected CRs, which only takes a few tens of $\omega_c^{-1}$.
While this estimate has been derived by assuming a CR current parallel to the magnetic field \citep[e.g.,][]{bell04,ab09}, it applies also to  shocks with any obliquity because the growth rate of the Bell instability is almost independent on the angle between the current and the initial field \citep{rs10, matthews+17}.

\begin{figure}
	\includegraphics[width=0.495\textwidth]{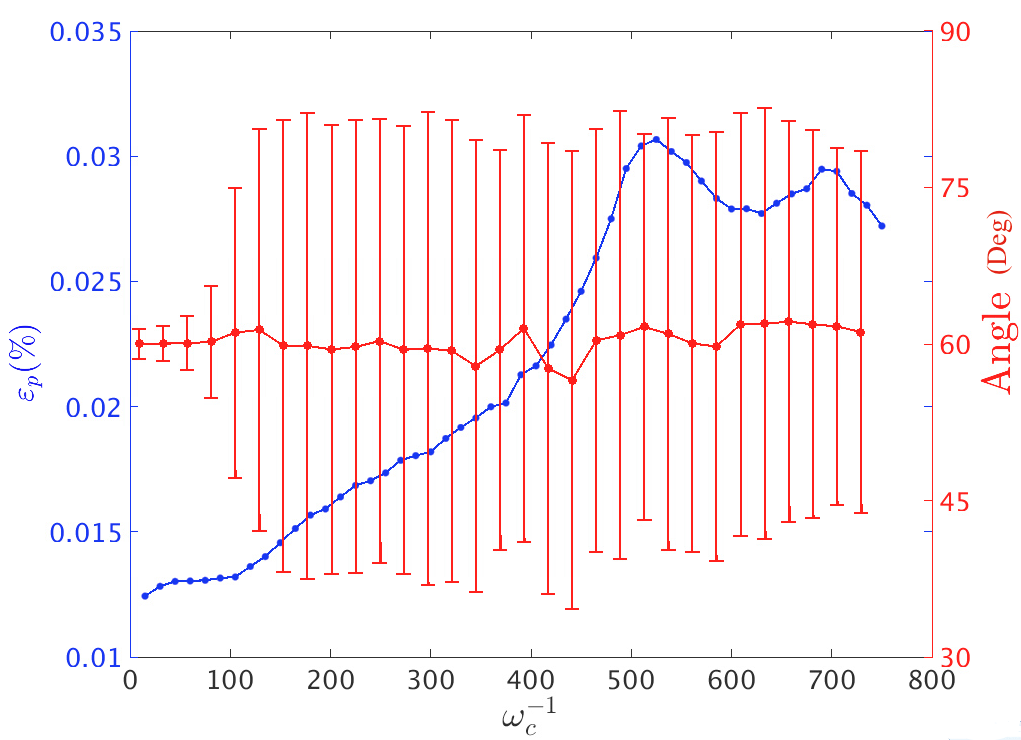}
	\includegraphics[width=0.495\linewidth ]{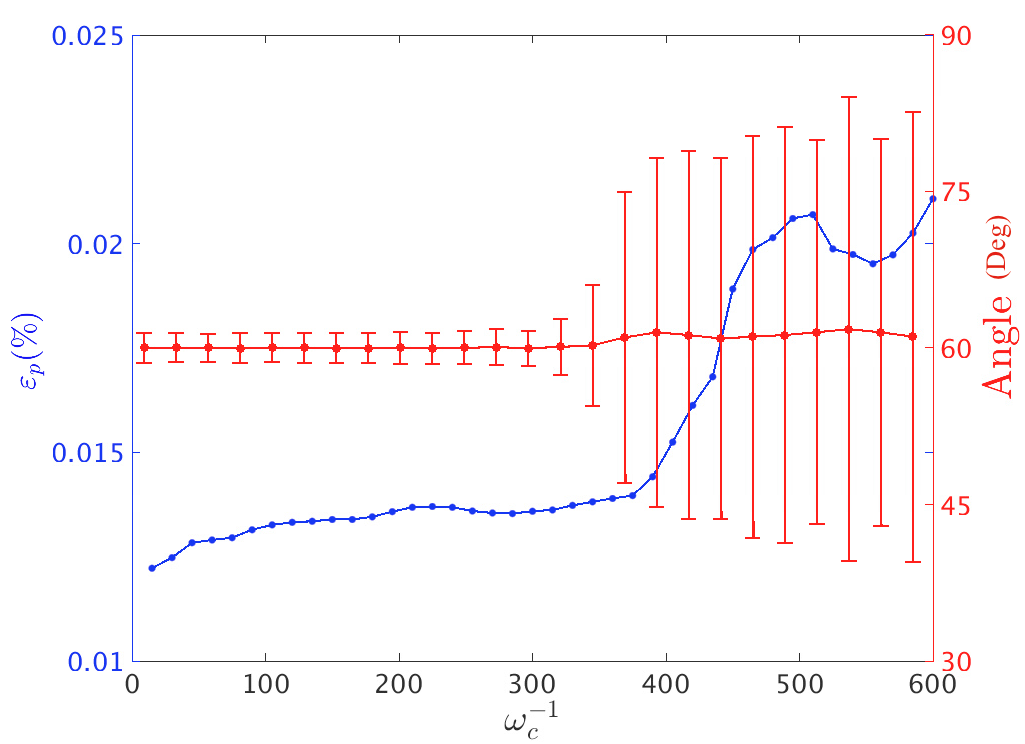}
	\caption{Time evolution of the proton acceleration efficiency $\epsp$ (left axes, blue) and of the effective shock inclination (right axes, red), for $\ncr = 0.01$ and $\ncr = 2\times 10^{-3}$ (left and right panel, respectively).
Error bars in the field inclination account for one standard deviation from the average, which is constant at the initial value of $\th= 60\deg$.
Note how $\epsp\lesssim 1\%$ until the onset of the Bell instability, which occurs later for the lower value of $\ncr$ (see Eq.~\ref{eq:tbell}).} \label{fig:effth}
\end{figure}

For comparison, we consider the case of a shock with the same benchmark parameters except that we put $\ncr=2\times 10^{-3}$ instead of $\ncr=0.01$;
therefore, the Bell instability is expected to develop a factor of five later in time according to Eq.\ \ref{eq:tbell}, also leading to a later trigger of proton injection into DSA.
Figure \ref{fig:effth} shows the time evolution of the acceleration efficiency $\epsp$ and of the effective inclination of the magnetic field at the shock for $\ncr = 0.01$ and $\ncr = 2\times 10^{-3}$.
The proton acceleration efficiency $\epsp\lesssim 1\%$ until $\tau_{\rm Bell}$, when the Bell instability starts to produce patches of quasi-parallel ($\th\lesssim 45\deg$) field.
The correlation between the onset of the Bell instability (in agreement with the theoretical prediction) and the increase in the proton acceleration efficiency demonstrates the crucial role of CR seeds in triggering proton DSA.

We conclude that, in the presence of energetic seeds, there is a typical timescale $\tau_{\rm Bell}$ determined by the current in reflected CRs after which the initial oblique magnetic field configuration is rearranged and thermal protons can be injected into DSA even at oblique shocks. 
In order to keep such a timescale within the range of accessibility of modern supercomputers, we use CR density much larger than those expected in the interstellar medium or in the solar wind, but we show in \S\ref{sec:SNR} that the extrapolation of $\tau_{\rm Bell}$ to astrophysical environments makes the effect relevant, for instance, for SNR shocks.

\section{Acceleration Efficiency: Dependence on Shock Inclination}\label{sec:eps}
Thus far, our results show that energetic CRs are re-accelerated in oblique shocks with $\th=60\deg$, and that in this case the proton acceleration efficiency is boosted to a few percent level. 
We investigate how CR re-acceleration and proton acceleration depend on the shock inclination by performing a series of 2D runs with $M = 30$ and different field inclinations from $0\deg$ to $80\deg$ (see table \ref{tab}).
Since at more oblique shocks a larger injection energy is required, we choose larger values of $\vcr$ to ensure that reflected CRs can be injected into DSRA; we also adjust $\ncr$ accordingly to keep the initial CR energy density $\lesssim 5\%$ of the proton one to ensure that the CRs are energetically subdominant. 

Note that, while the spectrum of Galactic CRs is dominated in number by trans-relativistic particles (see \S\ref{sec:SNR}), it is possible -- e.g., in heliospheric shocks -- to have seeds with spectra steep enough that most particles have non-relativistic energies $\vcr\gtrsim \vsh$.
Finally, this set of simulations also validates the model of \cite{injection} for the minimum injection needed to be injected into DSA.

\begin{table}
	\centering
	\begin{tabular} {ccccc}	
	\hline
	 Shock Inclination  & $\Delta t$ $(\omega_{c}^{-1})$ & $[L_x,L_y]$  $(c/\omega_p)$ & $\vcr$ $(v_A)$ & $\ncr (n_p)$  \\
	 \hline
	0$\deg$ & 0.0015 & [20,000, 200] & 50 & 0.01 \\ 
	\hline 
	30$\deg$ &  0.0015 & [20,000, 200] & 50 & 0.01 \\ 
	\hline 
	 45$\deg$  & 0.0015 & [20,000, 200] & 50 & 0.01 \\ 
	 \hline
	50$\deg$  & 0.0015 & [40,000, 200] & 50 & 0.01 \\ 
	 \hline
	60$\deg$ & 0.0015 & [40,000, 300] & 50 & 0.01 \\ 
	\hline
	70$\deg$ &  0.001 & [30,000, 200] & 90 & 0.006\\ 
	\hline 
	80$\deg$  & 0.0005 & [30,000, 200] & 200 & 0.0013\\
	\hline
	\end{tabular}
	\caption{\label{tab} Parameters for the 2D simulations of \S\ref{sec:eps}. All the shocks have $M$ = 30. } 
\end{table}

\subsection{CR Re-acceleration Efficiency}
\begin{figure}{}
\centering              
\includegraphics[trim=20px 50px 20px 320px, clip, width= 0.8 \textwidth]{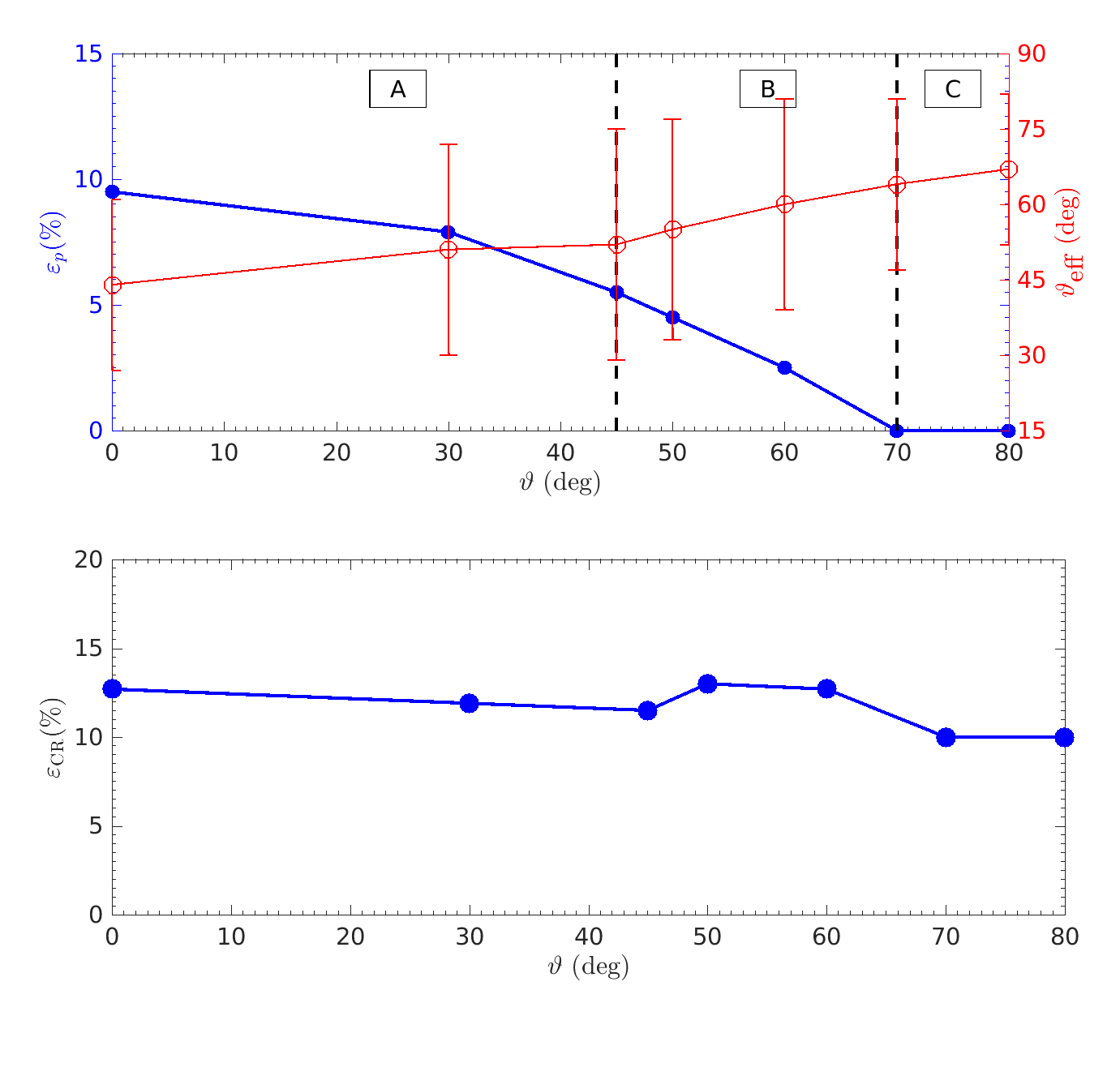}
\caption{CR re-acceleration efficiency $\epscr$ as a function of the shock inclination at $M = 30$ (see table \ref{tab} for the run parameters).
The absolute value of $\epscr$ has no intrinsic physical meaning because it scales linearly with $\ncr\vcr^2$, but the fact that CR DSRA efficiency is nearly independent of the shock inclination is a general result.}\label{fig:epscr}
\end{figure}

In addition to the proton acceleration efficiency $\epsp$, defined as the fraction of the post-shock energy density in ions with $E > 10\esh$, we introduce the CR re-acceleration efficiency $\epscr$ defined as the ratio of the total energy in re-accelerated CR seeds to the total energy in the downstream (which is dominated by the thermal protons).
Note that   $\epscr$ scales linearly with $\ncr\vcr^2$, which is typically much smaller than $n_p\vsh^2$ in realistic environments.
For our benchmark case ($\ncr=0.01$, $\vcr=50 v_A$, $\th=60\deg$) we find that the post-shock energy ratio between CRs and thermal protons is about $12\%$, a factor of $\sim 4$ more than far upstream.
Such a factor of 4 can be interpreted by considering that for $\vcr\gtrsim\vsh$ the downstream seed density increases by the shock compression ratio $\sim 4$, while their velocity remains roughly constant across the shock.

It is worth stressing that the re-acceleration efficiency is intrinsically limited by the number of seeds available: for spectra not flatter than $p^{-4}$, the maximum fraction of the shock kinetic energy that can be channeled in re-accelerated particles is not arbitrary and depends on $\ncr$. 

Figure \ref{fig:epscr} shows that the CR re-acceleration efficiency $\epscr$ does not depend greatly on the inclination angle for the runs with parameters in table \ref{tab}, being always between 10\% and 12\%.
The absolute values of $\epscr$ are rescaled to their values at $\ncr = 0.01$ and $\vcr = 50 v_A$ to allow for comparison between runs with different $\vcr$ and $\ncr$.

\subsection{Ion Acceleration Efficiency}
\begin{figure}{}
\centering              
\includegraphics[trim=20px 340px 20px 0px, clip, width=  \textwidth]{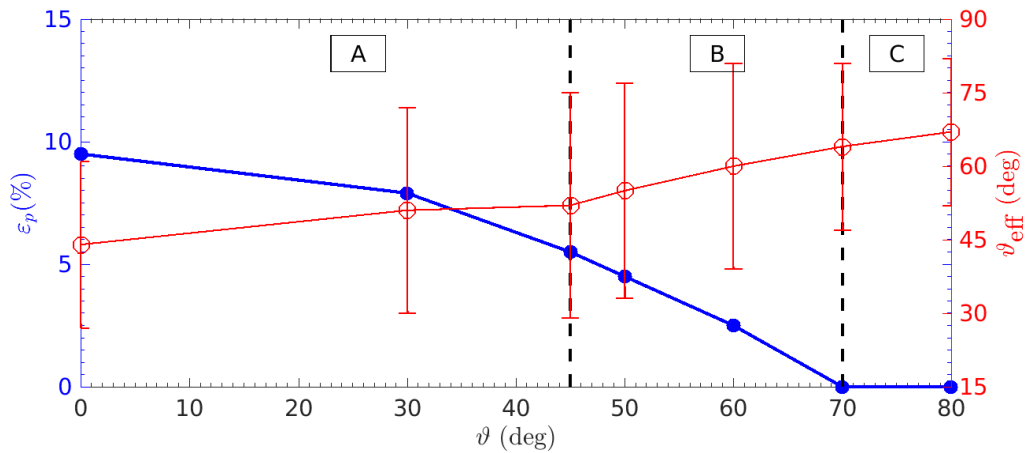}
\caption{Ion acceleration efficiency $\epsp$ as a function of the shock inclination at $M = 30$ (left axis, blue), along with the average upstream field inclination after the onset of the Bell instability (right axis, red); error bars indicate the standard deviation from the average field inclination.
The filling fraction of quasi-parallel regions decreases with increasing $\th$ and vanishes for $\th\gtrsim 70\deg$.
We distinguish three regimes. 
A: $\th \leq 45\deg$, where proton DSA is efficient regardless of the presence of CR seeds; 
B: $ 45\deg\lesssim \th \lesssim 60\deg$, where CR DSRA boosts the proton DSA efficiency; 
C: $ \th \geq 70\deg$, where even in the presence of CRs, ion DSA is absent.}\label{fig:epsp}
\end{figure}

Figure \ref{fig:epsp} illustrates the ion acceleration efficiency $\epsp$ for different shock inclinations in the presence of CR seeds (blue line). 
With respect to the case without CRs \citep[figure 3 of][]{DSA}, we identify three regimes characterized by the effectiveness of the CR-driven Bell instability in producing quasi-parallel regions in front of the shock. 
The red line in  figure \ref{fig:epsp} shows the effective shock inclination after $\tau_{\rm Bell}$ in each run. 


\begin{itemize}
\item Regime A, $\th \leq 45\deg$: 
protons can effectively be injected from the thermal bath and diffuse in the magnetic turbulence created by self-generated streaming instabilities \citep{DSA,MFA}.
The current in re-accelerated CRs increases the proton current only by  $\sim20\%$ for $\ncr=0.01$.
In astrophysical environments, where typically  $\ncr\ll 0.01$, we expect the proton current to vastly dominate the CR current and the overall shock acceleration efficiency not to depend on the presence of seeds.

\item Regime B, $50\deg\lesssim \th \lesssim 60\deg$: 
for these inclinations, the proton acceleration efficiency may be larger when seeds are present, because their re-acceleration provides a minimum current in the upstream.
Since the fraction of reflected protons (with velocity $\sim 2\vsh$) is not strictly zero, but drops exponentially with $\th$, the rearrangement of the magnetic field inclination is expected to happen after a timescale determined by the largest of the two currents, eventually triggering a more effective proton injection into DSA. 

\item Regime C, $\th \geq 70\deg$:
the fraction of reflected protons drops below $10^{-6}$, while there is still a reflected CR current. 
In this regime, however, the upstream magnetic field inclination cannot be rearranged to create quasi-parallel regions even if the Bell instability enters its non-linear stage (see the deviation from the average inclination in figure \ref{fig:epsp}).
For such quasi-perpendicular shocks injection of thermal protons is always strongly suppressed, and all the non-thermal activity depends on the presence and re-acceleration of seeds.
\end{itemize}

\subsection{Quasi-Perpendicular Shocks\label{sec:qperp}}
\begin{figure}{}
\centering
\includegraphics[trim=0px 30px 0px 20px, clip, width= .9 \textwidth]{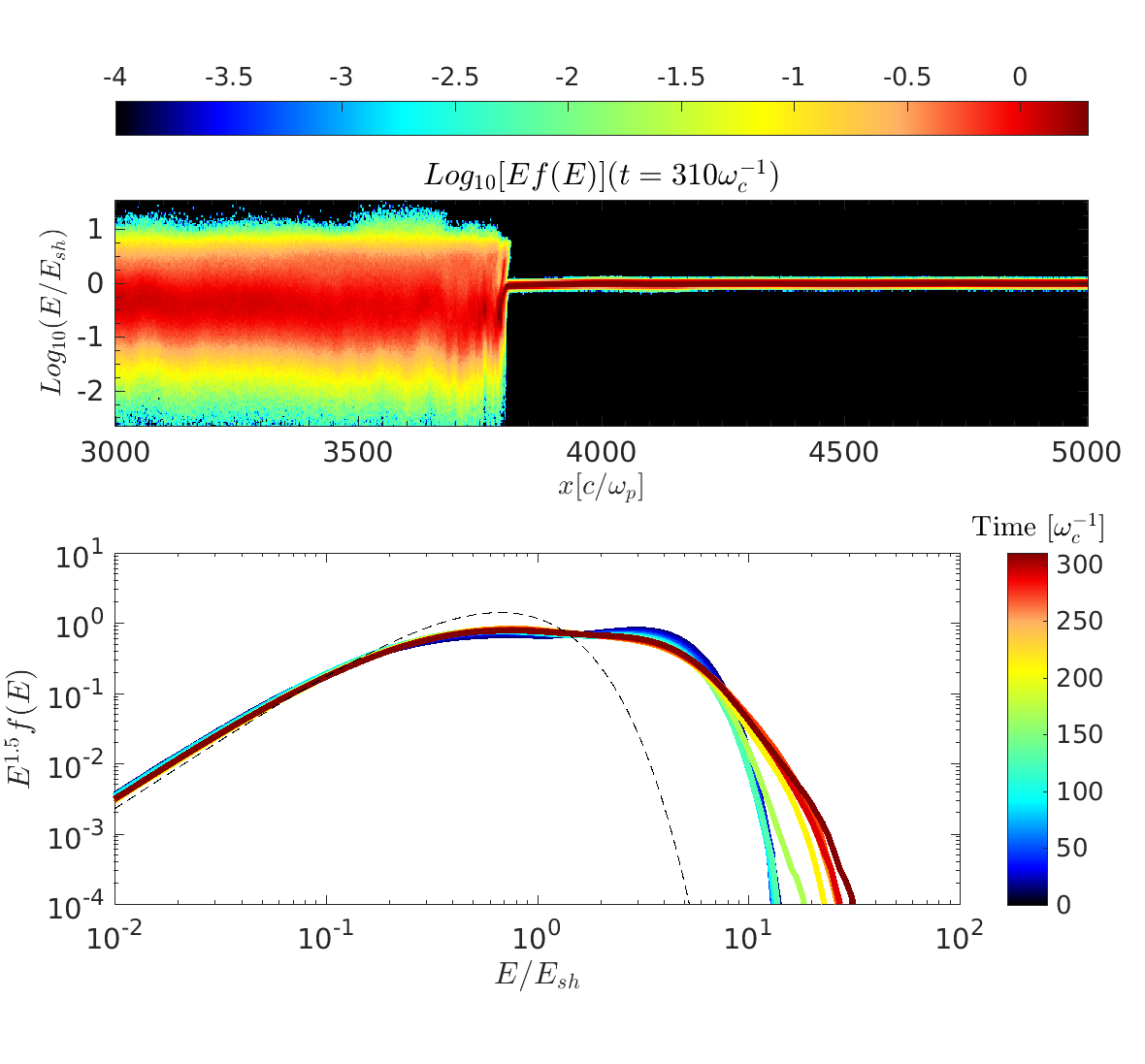}
\caption{Top panel: late-time proton energy phase space for $\th = 80\deg$. 
Bottom panel: time evolution of the downstream proton spectrum; the dashed line corresponds to the thermal distribution. 
Note that the maximum energy and the fraction of non-thermal ions grows with time after the onset of the Bell instability at $\tau_{\rm Bell}\approx 100\omega_c^{-1}$, but there are no energetic protons in the upstream, so DSA is ruled out as the acceleration process. }\label{fig:80ion}
\end{figure}

\begin{figure}{}
\centering
\includegraphics[width= \textwidth]{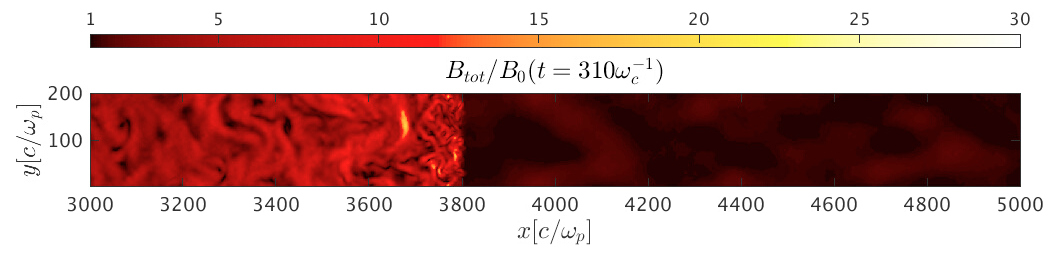}
\caption{ Magnetic field amplitude map around the quasi-perpendicular shock at $t = 310 \omega_c^{-1}$,  corresponding to the phase space plot in figure \ref{fig:80ion}. 
Note the non-linear upstream field amplification characteristic of the Bell instability driven by re-accelerated CRs and the turbulent downstream medium, which peaks behind the shock and decreases for $x\lesssim 3700c/\omega_p$, where non-thermal protons with $E\gtrsim 10\esh$ appear (see figure \ref{fig:80ion}).}\label{fig:80bf}
\end{figure}

\begin{figure}{}
\centering
\includegraphics[trim=0px 50px 0px 20px, clip, width= .9 \textwidth]{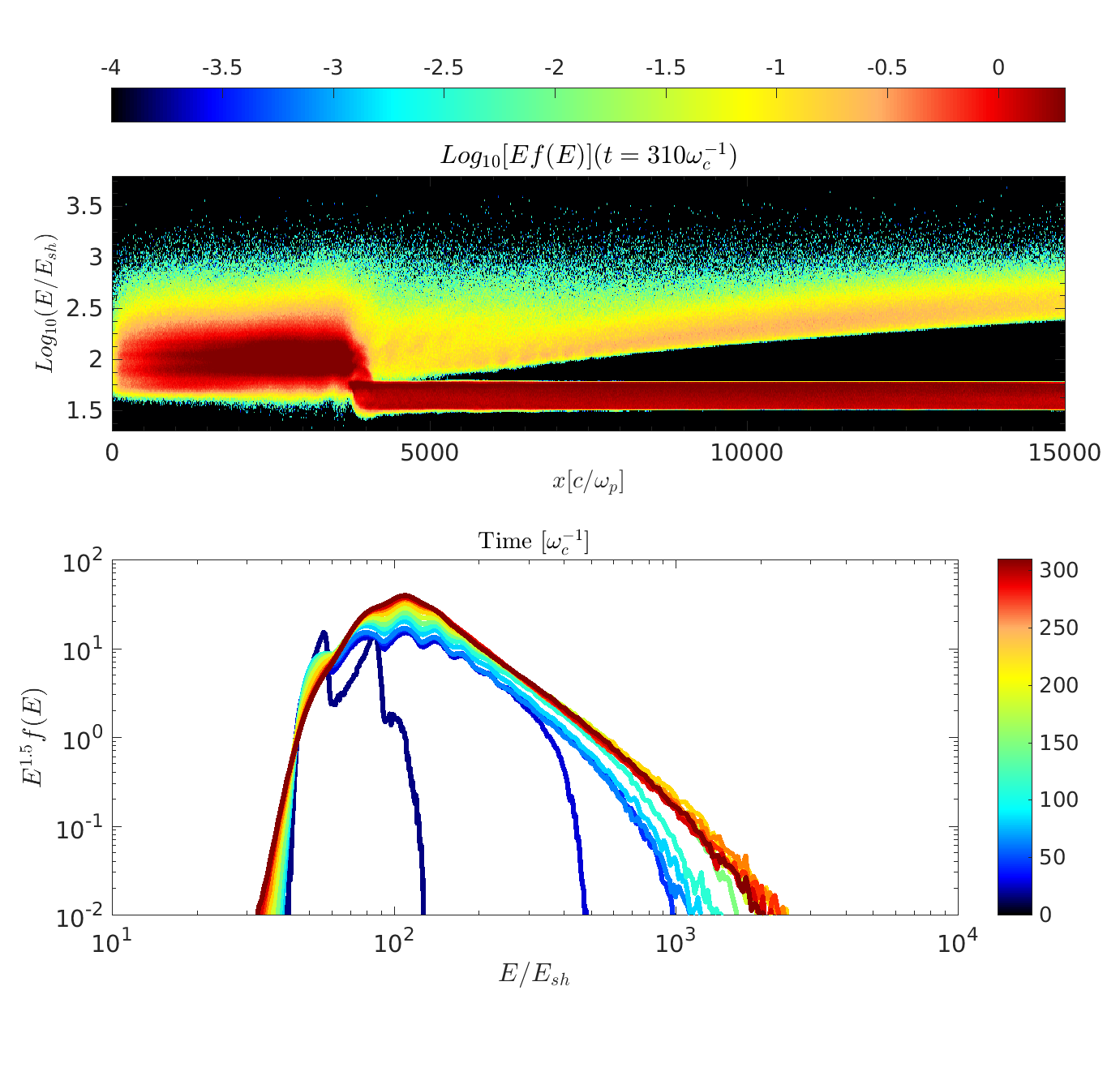}
\caption{As in figure \ref{fig:80ion}, but for CR seeds instead of protons. 
In this case there is a population of high-energy CRs escaping from the shock (top panel). 
Seeds are re-accelerated and form a power-law distribution that flattens with time and converges to $f(E)\propto E^{-4}$, significantly steeper than the DSA prediction, likely because of the larger fraction of particles that are removed by the acceleration process by being swept downstream.}\label{fig:80cr}
\end{figure}

Since our hybrid code is non-relativistic and the speed of light $c$ is effectively infinite, we cannot study superluminal shocks, i.e., configurations in which the velocity that a particle would need to overrun the shock by moving along the upstream magnetic exceeds $c$.
For non-relativistic shocks this regime is confined to almost perpendicular inclinations $\th'\geq \arccos (v'_{\rm sh}/c)$, where primed quantities are measured in the upstream frame.
In general, we do not expect any non-thermal activity for superluminal shocks \citep[see, e.g.,][]{ss09}.

Let us now consider in detail the run with $\th=80\deg$ in table \ref{tab}, which is representative of quasi-perpendicular shock configurations.
Figure \ref{fig:80ion} and figure \ref{fig:80cr} show the phase space and the time evolution of the downstream spectrum of protons and CRs, respectively, for such a quasi-perpendicular shock.
The proton spectrum exhibits the characteristic supra-thermal bump found in simulations without seeds CRs \citep{DSA}, but only at early times. 
At later times, after the CR-driven Bell instability develops, both the maximum energy of the ion spectrum and the fraction of non thermal protons with $E\gtrsim 10\esh$ grow. 
However, the top panel of figure \ref{fig:80ion} shows no energetic protons diffusing in front of the shock, so DSA cannot be responsible for such an energization. 
 The presence of CR-driven magnetic turbulence (figure \ref{fig:80bf}) provides an extra source of energy available to post-shock protons.
A comparison between figure \ref{fig:80ion} and \ref{fig:80bf} shows that the turbulent magnetic field peaks behind the shock (at $x\lesssim 3800c/\omega_p$) and dissipates for $x\lesssim 3700 c/\omega_p$, exactly where non-thermal protons with $E\gtrsim 10\esh$ appear.
Such a correlation suggests that  protons are energized either via second-order Fermi acceleration or via magnetic reconnection.
To our knowledge, this is the first time that such kind of acceleration for quasi-perpendicular shocks is reported in the literature;
a more detailed analysis including particle tracking and a thorough scan of the parameter space in $\th$ and $\vcr$ would be needed to fully characterize this acceleration mechanism, but they goes beyond the scope of this paper. 

Figure \ref{fig:80cr} shows that, unlike protons, energetic ($E\gtrsim 300\esh$) CRs can escape upstream (top panel) and be accelerated by being scattered back and forth around the shock.
The CR spectrum quickly develops a non-thermal tail, whose extent increases with time and whose slope converges to $f(E)\propto E^{-4}$, significantly steeper than the standard DSA prediction.
Since power-law distributions arise from the balance between acceleration rate and escape \citep{bell78a}, a possible explanation for such a steep spectrum is that the quasi-perpendicular shock geometry tends to trap and advect away from the shock a fraction of diffusing particles larger than at lower-inclination shocks. 
This effect, which involves higher-order terms in the anisotropy expansion of the CR distribution, has been studied, e.g., by \cite{bell+11}, but a direct comparison with such a formalism goes beyond the goal of this paper.
 
We can summarize the analysis of quasi-perpendicular shocks by remarking that, in the presence of CR seeds that can be re-accelerated and drive the Bell instability, two new acceleration features appear.
First, thermal protons can be accelerated in the downstream beyond the limit imposed by SDA, likely via either magnetic reconnection or second-order Fermi acceleration in the self-generated magnetic turbulence.
Second, CR DSRA leads to spectra significantly steeper than the standard prediction that hinges on isotropic particle distributions.

\section{A Universal Current in Reflected CRs\label{sec:jcr}} 
\begin{figure}{}
\centering
\includegraphics[trim=20px 40px 20px 234px, clip, width= .850 \textwidth]{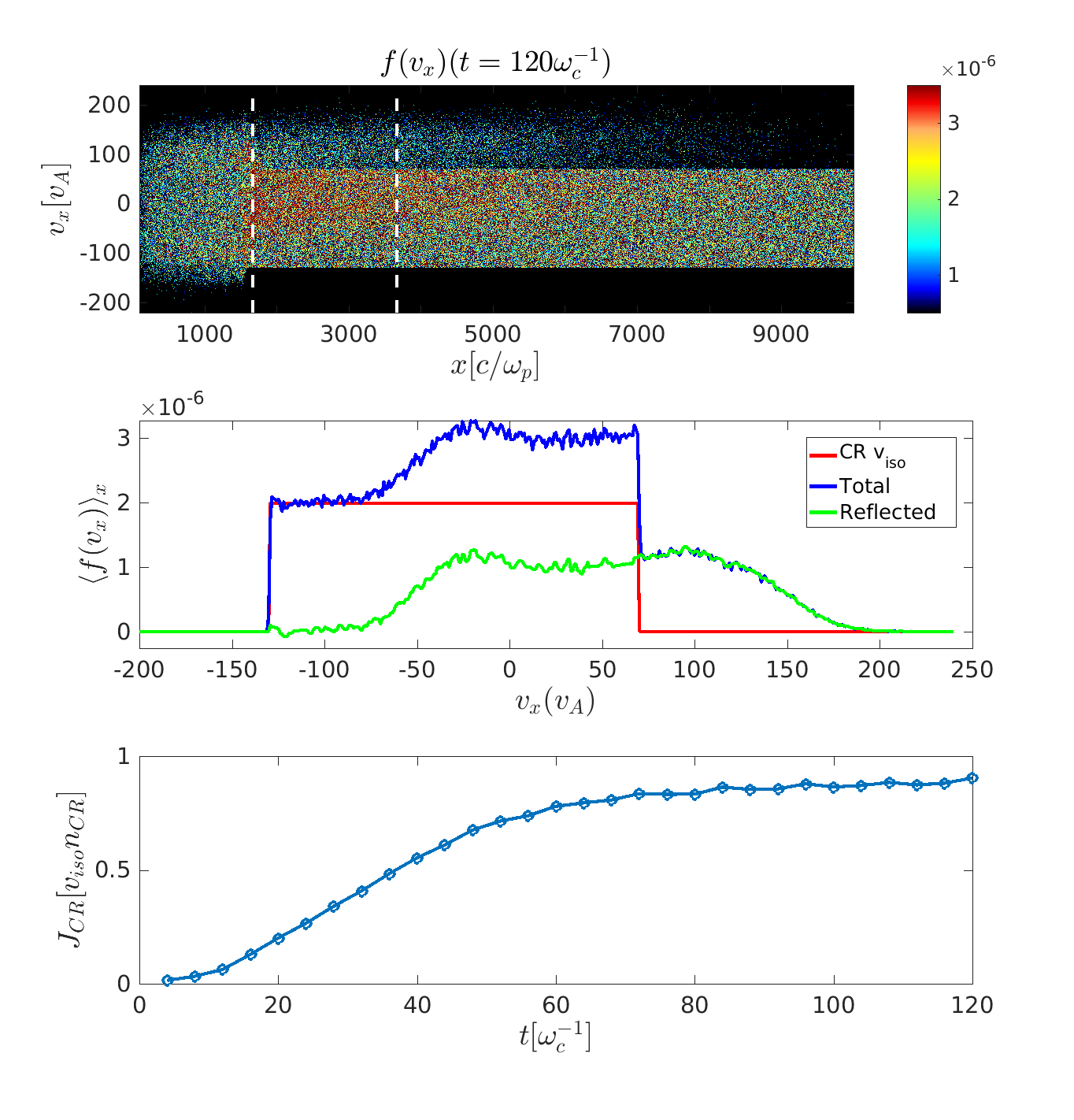}
\caption{
Top panel: velocity distribution $f(v_x)$ for the CR species, integrated in a region $\Delta x=2000 c/\omega_p$ immediately upstream of the shock at $t=120\omega_c^{-1}$. The time is chosen such that CR seeds have already been reflected but not effectively scattered, yet;  the results do not depend on the particular choice of $\Delta x$.
The distribution of reflected CRs (green line) is obtained as the difference between the total one (blue) and the initial isotropic one (red).
Bottom panel: time evolution of the reflected CR current, calculated as the integral over $v_x$ of the CR distribution above, which saturates to $\jcr\sim e\ncr\vsh$ after $\sim 60\omega_c^{-1}$. \label{fig:currcalc}}
\end{figure}

We have already outlined the crucial role played by the Bell instability generated by the reflected CR current $\jcr$, which we want to characterize in terms of the initial CR density and velocity, and of the shock parameters such as $M$ and $\th$. 
In other words, we now calculate the reflectivity of the shock for impinging CRs, in terms of both fraction of reflected CRs and reflected current $\jcr = \chi e\ncr\vsh$.  

For such an analysis we use periodic left and right boundary conditions for the CRs to ensure that an isotropic CR distribution velocity distribution impinges on the shock even at early times, when the shock is still forming. 
With open boundary conditions, in fact, CRs can gyrate out of the left boundary and leave without replenishing the supply of positive velocity particles ahead of the shock, breaking the CR velocity isotropy in front of the shock. 
Periodic left and right boundary conditions for the CRs circumvent this problem as the flow of positive-velocity CRs from the right boundary ensures that the pre-shock CR distribution is indeed isotropic since the very beginning. 
Once the shock moves away from the wall more than a few CR gyroradii, both open and periodic boundary conditions become equivalent. 
After this transient, $\jcr$ achieves a value that remains constant until $\tau_{\rm Bell}$, when non-linear perturbations start scattering CRs in pitch angle. 
We choose a low CR number density of $\ncr =  4\times 10^{-4}$ to have time to measure the saturation of the shock reflectivity before the onset of non-linear phenomena.

The current $\jcr$ is directed along the positive $x-$axis and can be calculated by looking at the $x-p_x$ phase space and integrating in $v_x$ the difference between the total distribution function $f(v_x)$ and the initial isotropic function, which is flat between $-\vcr$ and $+\vcr$ (we use also $v_{\rm iso}\equiv\vcr$). 
Figure \ref{fig:currcalc} shows the results of such a calculation for a case with $M = 30$, $\th = 60\deg$, and $\vcr = 100 v_A$. 
From the middle panel we see that the distribution of reflected CRs, $f_{\rm CR}^{\rm r}$ (green line), peaks slightly below $+\vcr$, with asymmetrical tails between $-\vcr$ and $+2\vcr$.
The bottom panel of figure \ref{fig:currcalc} shows that the CR current saturates for $t\gtrsim 50\omega_c^{-1}$, much earlier than the onset of the Bell instability (for these parameters, $\tau_{\rm Bell}\gtrsim 10^3\omega_c^{-1} $, see Eq.\ \ref{eq:tbell}).
From the plots in figure \ref{fig:currcalc}, we also determine the fraction of reflected CRs $\eta$, defined as
\begin{equation}
\eta \equiv \frac{\int f_{\rm CR}^{\rm r}(v_x) dv_x }{\int f^{\rm iso}_{\rm CR}(v_x) dv_x} 
\end{equation}
and the average velocity of the reflected CRs
\begin{equation}
v_{\rm u}\equiv\langle v_x \rangle \equiv \frac{\jcr}{\eta e\ncr}. 
\end{equation}

\begin{figure}{}
\centering
\includegraphics[width=  \textwidth]{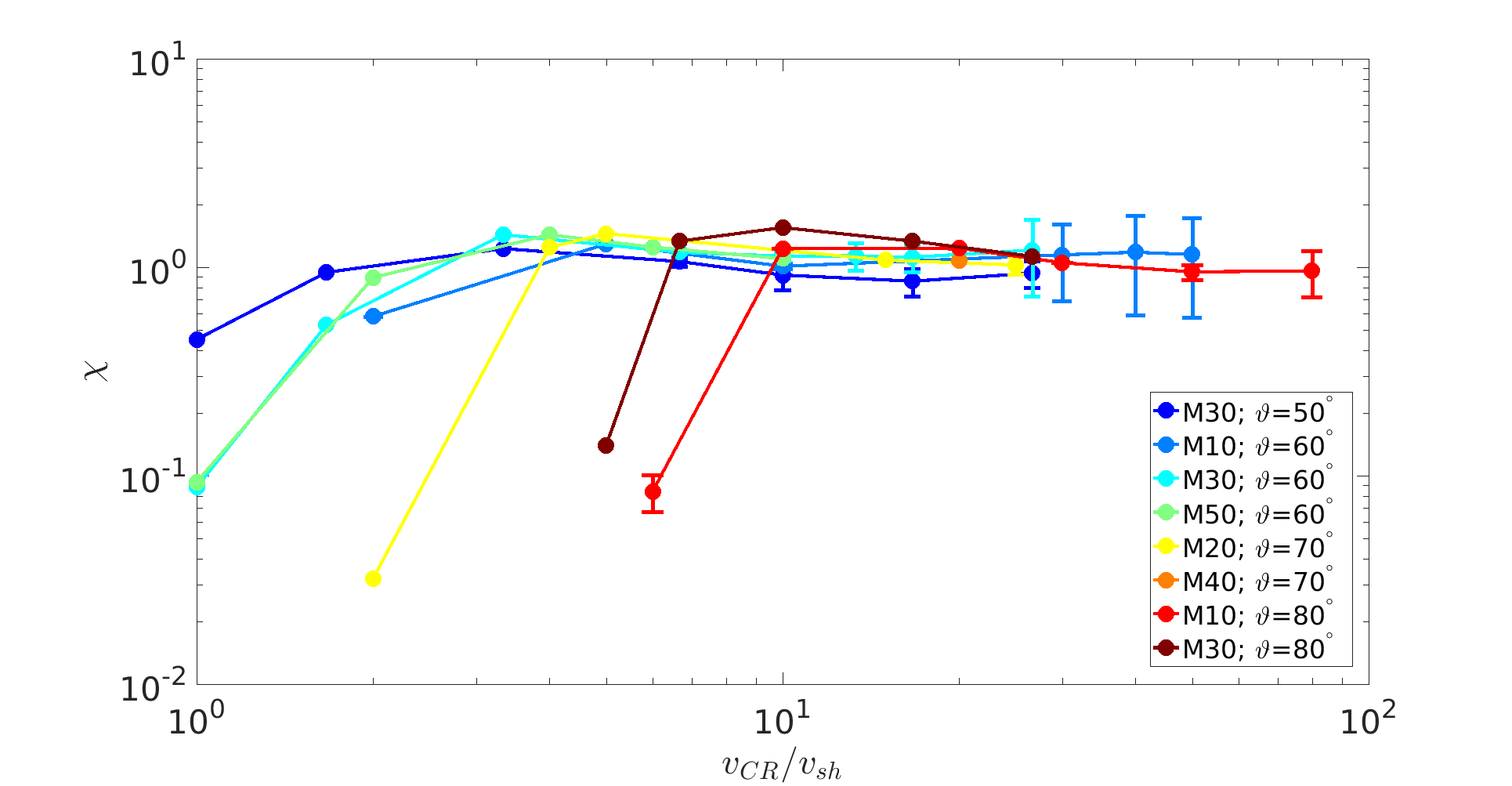}
\caption{Current in reflected CRs as a function of $\vcr/\vsh$ for shocks with different Mach numbers and field inclinations, as in the legend. 
For $\vcr \gg\vsh$, the reflected current $\jcr\simeq e\ncr\vsh$, independent of $M$ and $\th$. 
For $\vcr$ less than a few times $\vsh$, $\jcr$ drops steeply, and the location of such a drop depends strongly on the field inclination, consistent with the expectations for supra-thermal particles \citep{injection}.} \label{fig:currf}
\end{figure}

\begin{figure}
\includegraphics[width=0.49\linewidth ]{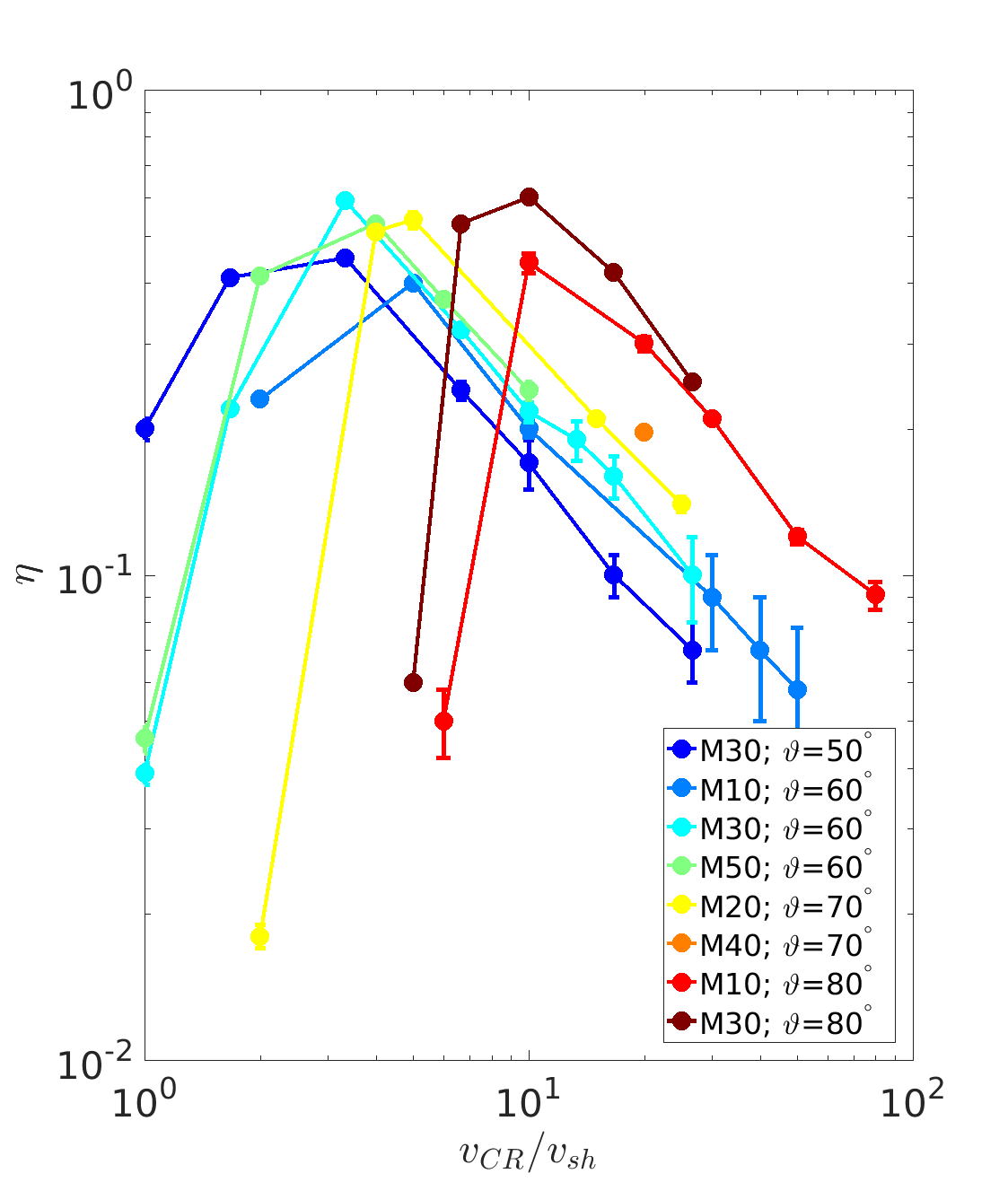}	\centering
\includegraphics[width=0.49\linewidth ]{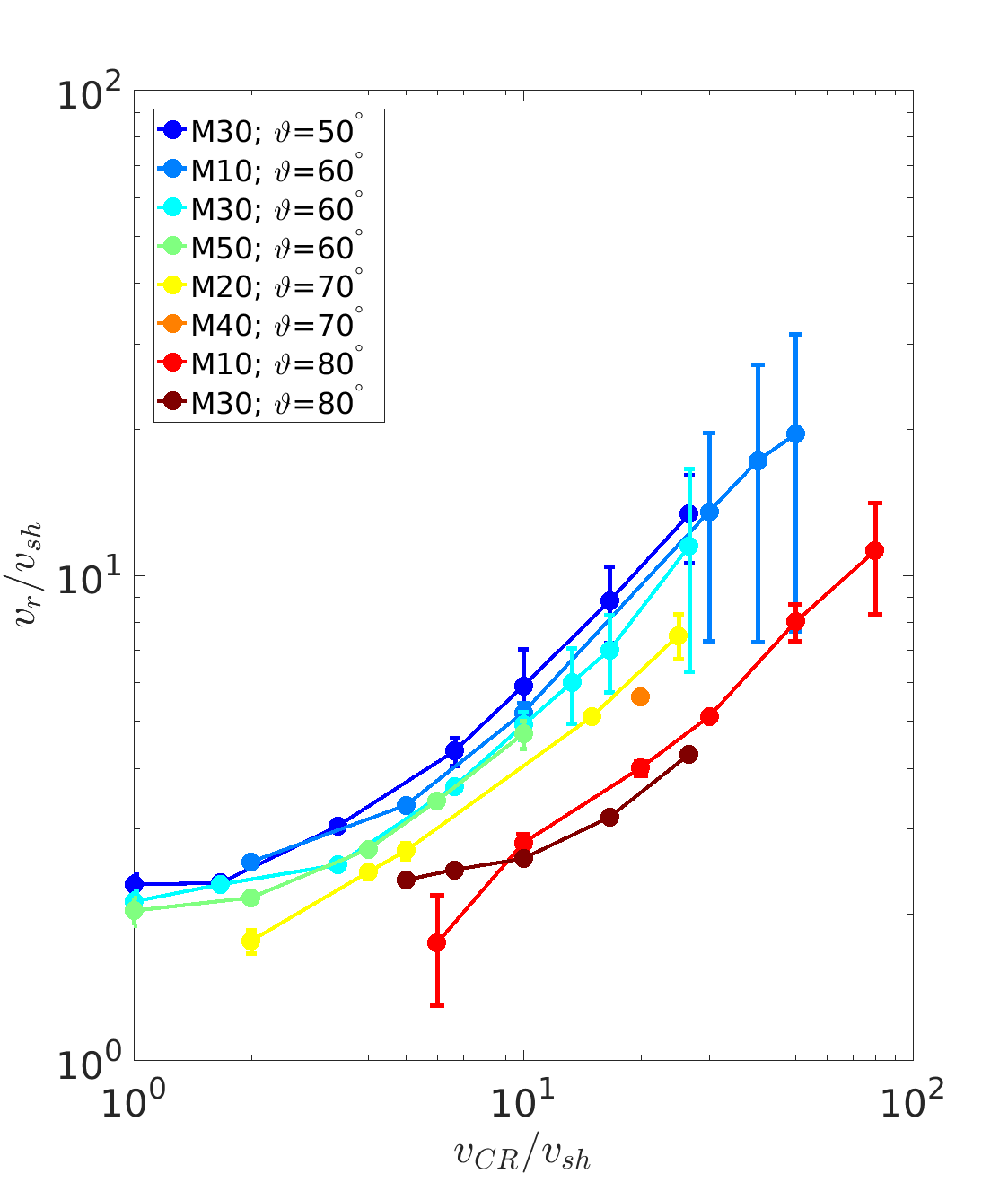}
\caption{Left panel: Fraction $\eta$ of CRs reflected at the shock.
$\eta$ increases for larger $\th$, and decreases steeply for $\vcr$ less than a few times $\vsh$ and linearly for $\vcr\gg\vsh$.
Right panel: average velocity of reflected CRs $v_{\rm r}$, which decreases with $\th$ and increases linearly for $\vcr\gtrsim\vsh$.
The combination of such trends returns the constant $\jcr$  in figure \ref{fig:currf}.} \label{fig:etav}
\end{figure}

Figure \ref{fig:currf} shows the normalized CR current 
\begin{equation}
    \chi\equiv\frac{e\int v_x f_{\rm CR}^{\rm r}(v_x) dv_x }{e \ncr\vsh}
\end{equation}
as a function of $\vcr$  for a range of Mach numbers and oblique to quasi-perpendicular  field inclinations.
Remarkably, for large values of $\vcr/\vsh$, $\chi$ approaches unity regardless of the shock properties. 
The very reason for such a universality can be understood by separating the contributions of $\eta$ and $v_{\rm r}$, as illustrated in figure \ref{fig:etav}.
The shock reflectivity (left panel) naturally drops if $\vcr \lesssim$ a few times $\vsh$, where the post-reflection velocity is smaller than the injection velocity, which strongly depends on the shock inclination \citep{injection}.
At the same time, $\eta$ decreases almost linearly for $\vcr\gg\vsh$ because very energetic particles with large rigidities tend not to see the shock discontinuity.  
The peak of reflectivity depends on the shock inclination, and increases with $\th$ at fixed $\vcr/\vsh$, because the more oblique shock effectively ``shrinks'' the CR gyroradius. 
The suppression of $\eta$ for $\vcr\gg\vsh$ is exactly compensated by the linear increase of $v_{\rm r}$, which is just proportional to $\vcr$ (right panel of figure \ref{fig:currf}).
Finally, at fixed $\vcr$, $v_{\rm r}$ decreases for large inclinations because CRs stream along the field lines, and a higher field inclination means a lower $x$ velocity. 
Such scalings hold also for relativistic seeds with $\vcr\approx c$, and for $\vsh$ up to $\sim c/2$;
for faster shocks, we expect the reflected current to be smaller because $v_{\rm r}$ cannot exceed $2\vsh$ as suggested in the right panel of figure \ref{fig:etav}.

In summary, for $\vcr\gg\vsh$ we expect a universal current due to reflected CRs, which has a very simple and elegant expression
\begin{equation}
\jcr\simeq e\ncr\vsh;\qquad \chi\simeq 1.
\end{equation}
Such a current seems to arise from a fine-tuned balance of the dependence of both $\eta$ and $v_{\rm r}$ on $\th$ and $\vcr/\vsh$, but the very reason for such an universality can be understood with the following argument.

The seeds distribution is initially isotropic in the upstream reference frame, but its interaction with the shock tends to drive it to isotropy \emph{in the shock frame}, as it is usually the case for particles whose gyroradius is much larger than the shock thickness. 
This is the standard assumption used to solve the CR transport equation \citep[e.g.,][]{bell78a}. 
If close to the shock, in the shock frame, there are $\ncr=\int {\rm d}\mathbf{v'} f'_{\rm CR}(\mathbf{v'})$ particles with  average flux $J'_{\rm CR}/e=\int {\rm d}v'_x v'_xf'_{\rm CR}(v'_x)=0$, boosting their distribution in the upstream frame produces a current $J_{\rm CR} = J'_{\rm CR}+e\ncr\vsh\approx e\ncr\vsh$, which corresponds exactly to the universal current.
However, note that, from the microphysical point of view, such a current is \emph{not} comprised by $\ncr$ CRs with velocity $\vsh$, but rather by fewer particles that overrun the shock because their velocities are larger than $\vsh$ (see figure \ref{fig:currcalc}).

\section{Application to Realistic Environments}\label{sec:SNR}
\subsection{SNRs in the Interstellar Medium}
We now make use of the fact that the reflected CR current is $\jcr \approx e \ncr\vsh$ to calculate the expected growth time of the Bell instability at SNR shocks due to the re-acceleration of Galactic CRs.

We start from the flux of Galactic CRs, $\phi(E)$, measured by the Voyager I spacecraft \citep{stone+13} and consider the non-relativistic part of such a flux, since it encompasses most of the particle number density. 
The transformation from flux to momentum distribution can be performed by using
\begin{equation}
4 \pi p^2 f(p)dp = \frac{4 \pi}{v(p)}\phi(E)dE. 
\end{equation}
Since $\phi(E)$ is roughly constant between 3 and 300 MeV \citep[see figure 3 of][]{stone+13}, we obtain that $f(p)\propto p^{-3}dE/dp \sim p^{-2}$, where we used that  $dE/dp \propto p\propto v$ for nonrelativistic particles.
Thus, the complete expression for the seed momentum distribution at low energies is 
\begin{equation}\label{eq:GCRs}
f_{\rm CR}(p) \approx {10^{-9}}{{\mathrm {cm}^{-3}}}\frac{1}{4\pi}p_0^{-3}\left (\frac{p}{p_0} \right)^{-2}. 
\end{equation}
where $p_0 \sim mc$, and we scaled the normalization to the typical CR energy density of 1eV/${\rm cm}^3$.
Such non-relativistic CR spectrum is rather hard, scaling as $p^{-2}$, and sets the level of seeds that can be reprocessed by SNR shocks. 
This scaling extends down to MeV protons, which have $v \sim\vsh$, and can be integrated to find $\ncr$ as
\begin{equation}\label{eq:ncr}
\ncr \approx \int_{p_{\rm min}}^{mc} 4 \pi p^2 \left [ \frac{10^{-9}}{{\rm cm}^{3}}\frac{1}{4\pi}p_0^{-3}\left (\frac{p}{p_0} \right)^{-2}\right] dp \approx {10^{-9}}{{\rm {cm}^{-3}}}\left (1- \frac{3\vsh}{c} \right),
\end{equation}
where we have put $p_{\rm min} \simeq 3m\vsh$ as the injection momentum. 
This choice is based on the fact that $\jcr$ is independent of $\vcr/\vsh$ above such a $p_{\rm min}$ and is consistent with our previous results \citep{DSA, injection}.

For $\vsh \sim 10^4$km s$^{-1}$, one obtains $\ncr \sim 9\times 10^{-10}  {\rm cm}^{-3}$, much smaller than the values used in the paper.
Finally, by using Eq.\ \ref{eq:tbell}, for $B_0 = 3 \mu$G and $n_p = 1 {\rm cm}^{-3}$ we get $\omega_c^{-1} \sim 35$s and $\tau_{\rm Bell}/\Xi  \sim 8.3 \times 10^{7} {\rm s} \sim 2.6 $yr. 
Even considering $\Xi\lesssim 10$, this timescale is much shorter than the typical dynamical SNR time of thousands of years, suggesting that the Bell instability has ample time to grow and amplify the upstream magnetic field to nonlinear levels. 

Note that if the CR current is relatively weak, i.e.,  
\begin{equation}\label{eq:resnonres}
    \ncr\vsh \lesssim 2 n_g v_A^2/c,
\end{equation}
left-handed, resonant modes are expected to grow at the same rate of Bell's right-handed, non-resonant modes \citep[see, e.g.,][]{ab09}.
The modes excited by the resonant streaming instability can quench the CR current as soon as  $\delta B/B_0\sim 1$ \citep[e.g.,][]{MFA}. 
However, such an amplification is already enough to change the effective shock inclination, so the phenomenology outlined above should still hold.
It is worth stressing that the streaming instability only requires CRs to be super-Alfv\'enic \citep[e.g.,][]{kp69}, which is always the case for reflected seeds.
In this sense, there is no threshold for the effect to be relevant, and the growth of CR-driven waves can only be limited by the possible presence of damping.
 In the interstellar medium CR seeds and magnetic fields are typically in equipartition, hence the total energy in reflected seeds, which is about four times the initial one, should generally be sufficient to generate non-linear fluctuations with $\delta B/B_0\gtrsim 1$.

\begin{figure}{}
\centering
\includegraphics[width= 0.7 \textwidth]{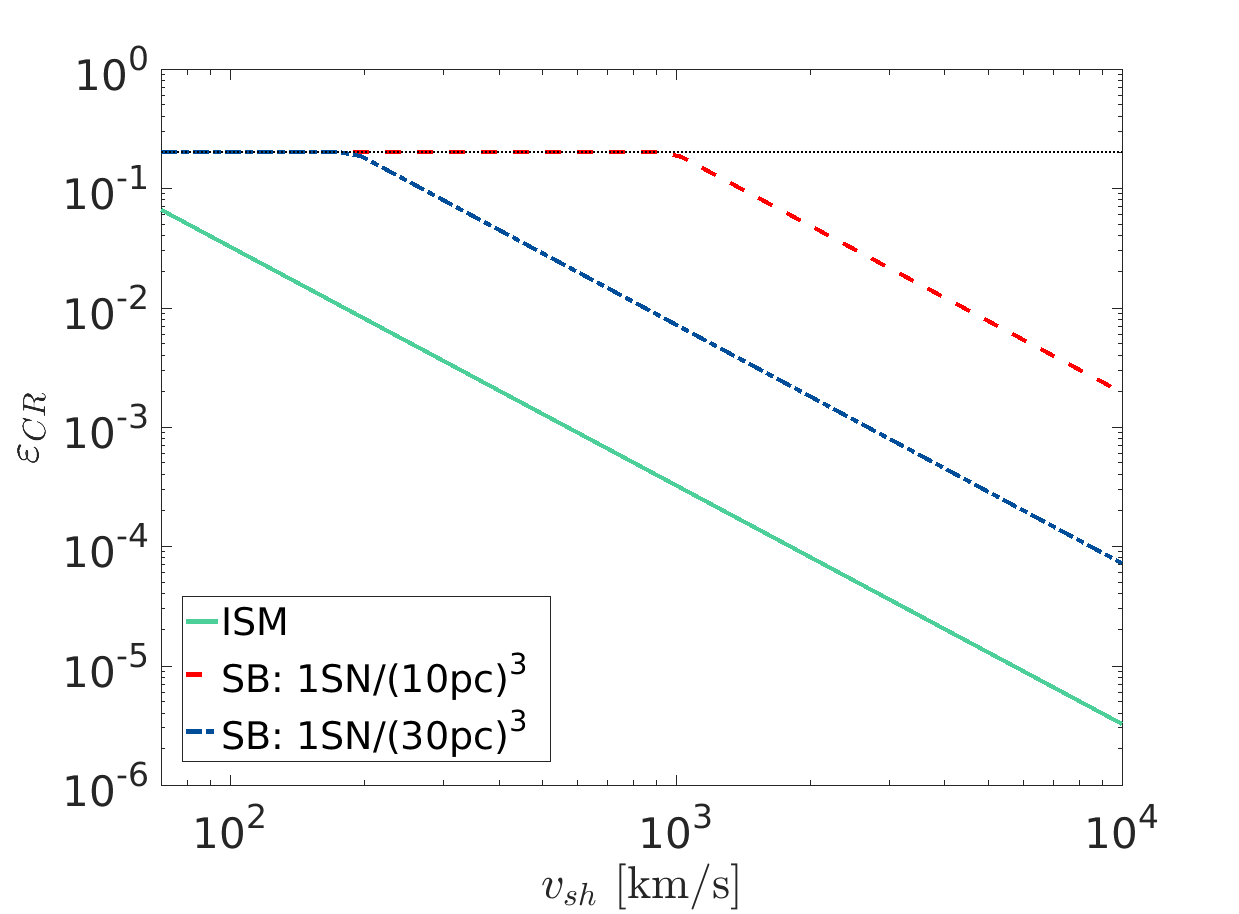}
\caption{Expected CR re-acceleration efficiency $\epscr$ as a function of the SNR shock velocity.
The solid line is for Galactic CRs, while dashed and dot-dashed lines illustrate superbubble cases, where the CR flux is enhanced due to multiple SN explosions (see Eq.~\ref{eq:ncrsb}).
For Galactic CRs $\epscr\gtrsim$ a few per cent for $\vsh \lesssim 300$ km s$^{-1}$, while in superbubbles re-acceleration should be important even for much faster shocks, possibly during the whole Sedov stage.
$\epscr$ is capped at $\sim 20\%$ based on the maximum efficiency obtained without seeds \citep{DSA}.} \label{fig:CRsh}
\end{figure}

Given the CR re-acceleration efficiency of $\sim 10\%$  for our reference parameters ($\ncr = 0.01$ and $\vcr = 50 v_A$, see Figure \ref{fig:epscr}), we can estimate the CR DSRA efficiency for a range of shock velocities simply by rescaling $\ncr$ and $\vcr$ to the actual values for Galactic CRs.
The solid curve in figure \ref{fig:CRsh} shows  $\epscr$ for typical interstellar values of $n_p = 1$ cm$^{-3}$, $\ncr = 10^{-9}$cm$^{-3}$, $v_A \sim 10$ km s$^{-1}$, $\vcr = c$, and $\vsh = 100 - 10,000$  km s$^{-1}$. 
The CR re-acceleration efficiency ranges from $\epscr\simeq 2\%$ for $\vsh = 100$ km s$^{-1}$ to $ \epscr \simeq 3\times10^{-6}$ for $\vsh = 10,000$ km s$^{-1}$, suggesting that DSRA may be important for isolated middle-age/old SNRs in the late-Sedov/radiative stages.

\subsection{SNRs in Superbubbles\label{sec:superb}}
Quite interestingly, there are active star-forming regions (often called \emph{superbubbles} or \emph{supershells}) where the SN rate is so high that SNRs effectively propagate in a medium that has recently been shocked, and therefore rich in energetic seed particles \citep[e.g.,][and references therein]{bykov14}.

Typically, superbubbles have radii tens to hundreds of pc and also contain O and B stars with powerful winds that can release up to $10^{51}$ erg of kinetic energy, comparable to a SN explosion. 
Stellar winds have velocities of hundreds to thousands of km s$^{-1}$, and can (re)accelerate particles as ordinary SNR blast waves.
Open star clusters (e.g., Westerlund 2) also host OB associations, young stars, and multiple SNRs, and show prominent non-thermal emission \citep[e.g.,][]{3FGL}. 

Let us estimate the content of seed particles in a superbubble by calculating the average CR density inside a homogeneous SNR of radius $R=R_{\rm pc}$ pc where about $E_{50}=10^{50}$ erg went into accelerated particles, corresponding to about 10\% of a typical SN explosion energy.
The CR spectrum should extend down to a few $m\vsh$ with a $p^{-4}$ power law, steeper than the spectrum of Galactic CRs in Eq.~\ref{eq:GCRs}.
However, since in this case the upstream energy density is also mostly in GeV particles, we still consider the number density in trans-relativistic seeds with $E\simeq mc^2$, which reads:
\begin{equation}
    n_{\rm CR,SN}\approx 6\times 10^{-4} \frac{E_{50}}{R_{\rm pc}^3}.
\end{equation}
If we introduce $ n_{\rm SN}= N_{\rm SN,wind}/R_{\rm pc}^3$ as the effective density of SNe (or stellar winds) per cubic pc, we can write the typical seed density in a superbubble as 
\begin{equation}\label{eq:ncrsb}
    n_{\rm CR,SB}\approx 6\times 10^{-4} E_{50} n_{\rm SN}.
\end{equation}
Here we have implicitly assumed that the typical delay between SN/wind events is smaller than the diffusive escape time of CRs from the SNR, which is very reasonable if the local diffusion coefficient is Bohm-like.

Dashed and dot-dashed lines in figure \ref{fig:CRsh} show the expected re-acceleration efficiency in a superbubble environment with $n_{\rm SN}=1/10^3$ and $n_{\rm SN}=1/30^3$, respectively.
For these reasonable SN/winds densities the availability of energetic seeds is enhanced by large factors ($\sim 600$ and $\sim 20$) with respect to the case of a SNR expanding in the interstellar medium (solid line).
As a result, seed DSRA \emph{alone} can lead to a very efficient production of non-thermal particles for shocks with quite large velocities $\vsh\lesssim 5\times 10^3$ km s$^{-1}$, and such an acceleration efficiency is independent of the shock inclination (figure \ref{fig:epscr}).
All the curves are arbitrarily truncated at a critical efficiency of $\epscr\sim 20\%$, beyond which the modification induced by non-thermal particles is expected to significantly smooth the shock transition, in turn suppressing particle injection \citep{DSA,injection}.

In summary, thanks to the abundance of seed particles, we expect shock (re)acceleration in superbubbles to be as efficient as possible for most of the SNR Sedov stage, regardless of the shock inclination. 
This has important implications also for the possibility of launching CR-driven winds from star-forming regions, which may play a crucial role in galaxy formation \citep[e.g.,][]{sb14, farber+17, no17, pfrommer+17}.

\subsection{SNRs Interacting with Molecular Clouds}
Another case in which CR re-acceleration is expected to be important is when SNR shocks encounter dense molecular clouds, as in W44 or IC443, which are prominent sources of hadronic $\gamma$-rays \citep{pionbump}.
As shown by different authors \citep[e.g.,][]{uchiyama+10,cardillo+16}, the observed $\gamma$-ray spectrum can be explained without invoking DSA of thermal protons but as simply due to the re-acceleration of the low-energy Galactic CRs that should be trapped inside the molecular clouds.
By assuming that the density of CRs is proportional to the gas density, in dense clouds one would infer $\ncr$ few orders of magnitude larger than the estimate in Eq.\ \ref{eq:ncr}.
Moreover, since these SNRs are quite old, their shock speeds, $\vsh\approx 200-500$km s$^{-1}$, fall exactly in the regime where DSRA is expected to be more efficient (figure \ref{fig:CRsh}).
The combination of these two factors suggests that in dense clouds the DSRA efficiency could easily be as large as 10--20\%.

\subsection{Heliospheric Shocks}
Seed re-acceleration is also expected at heliospheric shocks, since the solar wind is often rich in energetic particles produced, for instance, in solar flares.
In these cases, the peculiar chemical composition observed in \emph{solar energetic particle} events \citep[e.g.,][]{mason+04,tylka+05} represents a powerful diagnostics for investigating the interplay between shock inclination, seed re-acceleration, and thermal particle acceleration.
We also stress the analogies between the efficient re-acceleration of energetic seeds and the preferential acceleration of ions with large mass/charge ratios \citep{AZ}, since both particles share the property of having gyroradii (much) larger than thermal protons.

In the solar wind, pre-shock distributions are often not Maxwellian but rather \emph{kappa-distributions} with a power-law-like tail at supra-thermal energies. 
Such supra-thermal particles can thereby act as seeds and be injected into DSA also for oblique shocks, differently from what happens for thermal particles. 
In general, the level of pre-existing seeds in heliospheric shocks varies greatly from event to event, so it is impossible to provide a universal estimate for their re-acceleration efficiency, as we did for SNRs and diffuse Galactic CRs.
Nevertheless, the criterion for ion injection defined in \cite{injection} and the seed phenomenology outlined here should suffice for interpreting a given heliospheric event once the far upstream conditions (shock strength and inclination, seed abundance and chemical composition) are measured.

The re-acceleration mechanism outlined here is relevant also for \emph{pickup ions} in heliospheric shocks \citep[e.g.,][]{zank+96,kallenbach+00,heerikhuisen+10}.
Pickup ions are created when neutral particles (the origin of which may be interstellar, lunar, cometary, or due to dust sputtering) are ionized and/or undergo charge exchange with solar wind protons.
They typically enter heliospheric shocks with non-Maxwellian distributions that exhibit high-velocity tails, thereby acting as the non-relativistic CR seeds considered in this work. 
In particular, pickup ions that are swept outward to the solar wind termination shock can be re-accelerated to multi-GeV energies and are usually referred to as \emph{anomalous CRs} \citep[see][for a review]{cs99}.

\subsection{Clusters of Galaxies}
Another astrophysical environments in which ion re-acceleration may play a crucial role are clusters of galaxies, where GeV particles are expected to be confined on cosmological timescales \citep[see, e.g.,][for a review of non-thermal activity in clusters]{bj14}. 
Intracluster shocks have velocities comparable to SNRs, but smaller sonic Mach numbers because of the higher temperature plasma; the Alfv\'enic Mach numbers, instead, are comparable to those of SNR shocks since the upstream field is of the order of $\sim 1$ $\mu$G. 
Since the current in reflected seeds only depends on the Alfv\'enic Mach number, we expect that the phenomenology outlined above should apply to intracluster shocks as well.

\section{On the Injection of Thermal Ions at Oblique Shocks}
\subsection{The Role of Pre-existing Turbulence}
Since one of the main effects of DSRA  is to generate  \emph{intrinsic} magnetic turbulence at shocks of any obliquity,  hence enabling the injection of thermal protons into DSA also at oblique shocks, it is natural to discuss the potential role of the \emph{extrinsic} turbulence usually present in astrophysical plasmas. 

Hybrid simulations have shown that -- if large ($\delta B/B_0$) Alfv\'enic turbulence is present upstream -- injection of thermal protons is possible also for shocks  that are perpendicular \emph{on average} \citep[e.g.,][]{giacalone05}.
In these simulations  the coherence length of the magnetic field, $L_c$,  with $\delta B(k_{\rm min}=2\pi/L_c)/B_0)\approx 1$, was chosen to be a factor of 10-100 larger than the gyroradius $r_L$ of supra-thermal particles, i.e., of particles with velocity a few times $\vsh$;
therefore, some supra-thermal protons, while drifting along the shock surface, may encounter a quasi-parallel patch that allows them to escape upstream, which would not be kinematically allowed in a perpendicular shock \citep[see, e.g.,][]{STG83,injection}.
In this case,  the fraction of injected particles \citep[$10^{-4}$ in][]{giacalone05} is much smaller than the 1$\%$ measured at quasi-parallel shocks  \citep[e.g.,][]{DSA}, and the spectrum of the accelerated particles is not consistent with the standard DSA prediction.

The importance of extrinsic turbulence for astrophysical shocks, however, may be quite limited.  
The typical coherence length of the magnetic field in the Galaxy is  $L_c\approx 100$ pc \citep[e.g.,][]{jf12,beck+16}, several order of magnitude larger than the gyro-scales of supra-thermal particles,   $r_L\approx  3\times 10^{12} (\vsh/c) B_{\mu \rm G} $ cm, where $ B_{\mu \rm G}$ is the magnetic field in $\mu$G and $\vsh/c\sim 10^{-2}-10^{-3}$ for SNR shocks.
If we consider  a Kolmogorov-like scaling for the spectrum of the magnetic turbulence, $\delta B(k)/B_0\propto k^{-5/6} $ (neglecting anisotropy and damping), the amplitude of the extrinsic interstellar turbulence is extremely small ($\delta B/B_0\lesssim 10^{-9}$) at the scales relevant for the injection of thermal protons.
Even accounting for the additional magnetic turbulence due to the streaming of diffuse Galactic CRs, whose amplitude is $\delta B/B_0\approx 10^{-3}$ at scales resonant with GeV particles with  $r_L\approx 3\times 10^{12}$ cm \citep[e.g.,][]{aloisio+15,zweibel17}, it is unlikely that suprathermal particles in SNR shocks can experience a change in the local direction of the magnetic field due to pre-existing turbulence. 

The situation may be different for heliospheric shocks, since the solar wind is rich in magnetic structures at  much smaller scales \citep[e.g.,][for a review]{alexandrova+14}, but the potential role of any extrinsic turbulence would need to be assessed on a case-by-case basis and  also account for the actual spectrum of the magnetic fluctuations, its anisotropy, and for possible inhomogeneities and intermittency.
Normalization, slope, and maximum energy of the spectrum of the thermal protons accelerated at turbulent quasi-perpendicular shocks  in general depends on all of these details.

Conversely, the amount of CR seeds in the interstellar medium is well constrained (see discussion above) and the instabilities that they drive naturally generate non-linear magnetic perturbations on scales just slightly larger than the gyroradii of supra-thermal particles, which is expected and shown (\S\ref{sec:feedback}) to inject and accelerate thermal protons with a few percent efficiency. 

Finally, it is possible that large-scale magnetic fluctuations could channel energized particles from quasi-parallel to quasi-perpendicular regions, eventually triggering a magnetic field reorientation conducive to the injection of thermal ions, but this has never been quantified with self-consistent kinetic simulations.
Nevertheless, the bilateral morphology of the non-thermal emission from SNRs such as SN1006 and G1.9+0.3 favors a scenario where  ion DSA is efficient only in quasi-parallel region, while electron acceleration (or re-acceleration) can occur regardless of the shock inclination  \citep{icrc15}.

\subsection{Test-particle and MHD-PIC Simulations of Oblique Shocks}
Hybrid and full-PIC simulations clearly show that, without seeds, oblique shocks in laminar plasmas cannot trigger and sustain DSA, the culprit being the exponential suppression of the number of thermal protons that can overrun shocks with larger and larger inclination \citep[][and references therein]{injection}.
If injection is provided either with seeds or through a local re-arrangement of the magnetic field, DSA can occur also at oblique shocks and accelerated particles can trigger magnetic field amplification \citep[DSA is expected to be even faster at oblique shocks, see, e.g.,][]{jokipii87}. 

The fraction of particles injected into DSA is regulated by the quasi-periodic reformation of the shock barrier and by phenomena that occur on the gyro-scales of  thermal ions. 
If such structures are not resolved, for instance in test-particle simulations or if  thermal particles are accounted for in the MHD approximation  \citep[e.g.][]{uHybrid,vanmarle+18}, it is impossible to quantify proton injection. 
In particular, the MHD background cannot reproduce the time-dependent shock \emph{overshoot}, which acts as a barrier that, on one hand, reflects incoming particles and, on the other hand, prevents the leakage of downstream thermal particles into the upstream; 
not capturing the overshoot generally leads to over-injection of post-shock thermal ions.
In a similar fashion, propagating test-particles in analytically prescribed or MHD-based electromagnetic fields in general does not reproduce either the correct fraction of injected particles, or its dependence on the shock inclination.
When the thermal plasma is not treated kinetically, the number and the phase space distribution of the injected particles are effectively free parameters.

Very recently \cite{vanmarle+18} put forward hybrid--MHD simulations \citep[\'a la][]{uHybrid} and claimed that, with an ad-hoc  injection prescription, some supra-thermal particles were able to overrun and oblique shocks, eventually triggering non-linear magnetic field amplification. 
They injected a fixed fraction $\eta=2\times 10^{-3}$ of particles with $v_{\rm inj}=3\vsh$, initializing them as isotropic just behind the shock; such a prescription is taken from the hybrid simulations of  \emph{quasi-parallel} shocks  by \cite{DSA}, but it does not  apply to oblique shocks.
When we ran our benchmark simulation ($M=30$, $\th=60\deg$) without seeds and with a larger transverse size of $L_y= 1000 c/\omega_p$, we found no sign of non-thermal activity or magnetic field amplification up to $t\sim 1000\omega_c^{-1}$  (dotted curve in figure \ref{fig:pspec}), thereby confirming that oblique shocks inject much fewer thermal protons than quasi-parallel shocks.

The overall phenomenology that arises from any simulation where particles are injected by hand may look quite similar to the one described in the present work, but  it is important non to mistake bona-fide seeds for spontaneously-injected thermal particles, which can be accounted for only in kinetic simulations.

\section{Conclusions}\label{sec:conc}
We have presented the first comprehensive set of hybrid simulations that addresses the re-acceleration of pre-existing energetic particles in non-relativistic collisionless shocks and its effects on the global shock dynamics, in particular on proton injection and acceleration.
Our findings are summarized here in the following.
\begin{itemize}
\item Seeds with sufficiently large energy (few times $\vsh$, depending on the shock inclination) are effectively reflected at the shock, creating a current that drives the streaming instability in the upstream medium.
Seeds can then be scattered back and forth across the shock, diffusing in the self-generated magnetic turbulence, and develop power-law tails via a process that we dub diffusive shock re-acceleration, DSRA (figure \ref{fig:spec}).
\item The streaming instability can occur either in the resonant or in the Bell regime, depending on the strength of the current (Eq.~\ref{eq:resnonres}), but the general result is that when the magnetic field amplification becomes non-linear, the effective shock inclination changes (figure \ref{fig:theta}).
At oblique shocks ($50\lesssim\th\lesssim 70\deg$), where proton injection is normally inhibited \citep{DSA,injection}, the field rearrangement creates quasi-parallel shock regions where thermal protons can also be injected into DSA (figure \ref{fig:theta}).
\item For quasi-perpendicular shocks ($\th\gtrsim 70\deg$), seeds can still drive Bell waves and undergo DSRA, but the injection of thermal protons is always suppressed (figure \ref{fig:epsp}).
However, in this regime two new phenomena appear: 
first, the protons are accelerated in the downstream thanks to the CR-driven turbulence (figure \ref{fig:80ion});
second, the seeds are accelerated via DSRA with a very steep spectrum ($\propto E^{-4}$, see figure \ref{fig:80cr}) which does not depend only on the compression ratio, only, violating the prediction of standard DSA.
\item For $\vcr\gg\vsh$, the current in reflected CRs is universal and reads $\jcr\simeq e\ncr\vsh$, independently of shock Mach number, inclination, and $\vcr$ (figure \ref{fig:currf}).
Simulations and theory (\S\ref{sec:jcr}) explain this in terms of conservation of the seed anisotropy at the shock in the limit in which the seed gyroradii are much larger than the shock thickness.
\item For SNR shocks propagating into the interstellar medium filled with Galactic CRs, the growth time of the Bell instability due solely to the universal current in reflected CRs is of order of few years only;
this means that a minimum level of magnetic field amplification at SNR shocks must be expected, regardless of the shock inclination.
\item For middle-age/old SNRs with $\vsh$ of a few hundred km s$^{-1}$, DSRA of Galactic CRs alone can yield a total acceleration efficiency of few per cent; such an efficiency becomes much larger if the shock propagates in regions (e.g., superbubbles) where the CR density is significantly enhanced (figure \ref{fig:CRsh}).
\end{itemize}

Finally, it is worth mentioning that, while in the presented hybrid simulations we have considered only seed protons, the same re-acceleration mechanisms should apply also to heavier ions, which may be important for explaining the secondary/primary ratios in Galactic CRs \citep[e.g.,][]{blasi17}, and for seed \emph{electrons}, which may have phenomenological implications for the multi-wavelength emission of middle-age SNRs, for the non-thermal emission from clusters of galaxies, and may be crucial for interpreting spacecraft observations of energetic electrons in heliospheric shocks. 

\appendix

\section{}\label{appA}
The main assumption of hybrid models with kinetic ions and fluid electrons is that the dynamic scales of interest are those of the ions, while the dynamics of the electrons can be neglected \citep[e.g.,][]{winske85,Lipatov02, winske+03}. This translates to neglecting the displacement current in Amp\`{e}re's Law,
\begin{equation}
{\nabla} \times {\bf B}=\frac{4\pi}{c} {\bf J}.
\label{eq:amperedisp}
\end{equation}
thus suppressing the propagation of electromagnetic waves traveling at the speed of light. Also, an MHD model is considered for the electrons, and quasi-neutrality is assumed. Differences between various hybrid approximations depend mainly on whether the effects of finite electron mass, resistivity, and electron pressure need to be included in the MHD equations. To derive the hybrid set of equations, we start from the non-relativistic Vlasov equation for the electrons,
\begin{equation}
\frac{\partial\,f_e}{\partial\,t}+{\bf v}_e\cdot{\bf \nabla} f_e-\frac{e}{m}\left({\bf E}+\frac{{\bf v}_e}{c}\times{\bf B}\right)\cdot{\bf \nabla}_{v_e}f_e=0
\label{eq:vlasov}
\end{equation}
where $f_e=f_e\left({\bf r},{\bf v}_e,t\right)$ is the electron distribution function. In the current version of \textit{dHybrid}, the effect of finite electron mass is not considered (it is typically important on the electron skin depth scale, which is not resolved), and there is no explicit resistivity.
However,  the discretization of the distribution function and the finite spatial resolution introduce some noise that might be seen as a numerical resistivity, which is kept under control by checking convergence of the results with the number of macro-particles per cell, $N_{\rm ppc}$.
For strongly non-linear problems such as shock simulations, the physical signals are typically much larger than such a numerical noise, and a few particles per cell can be used\footnote{When species with different mass to charge ratio are present, we typically use $N_{\rm ppc}\sim\mathcal O(100)$ for the dynamically-dominant species to reduce the  numerical noise due to finite phase-space resolution \citep[see, e.g.,][]{AZ}}. 
In the $m \rightarrow 0$ limit of Eq. (\ref{eq:vlasov}), considering  Eq. (\ref{eq:amperedisp}), and an arbitrary number $N_{\rm sp}$ of ion species described as kinetic particles, the electric field is deduced to be:
\begin{equation}
{\bf E}=-\frac{{\bf V}_i}{c}\times{\bf B}+\frac{1}{4\pi n\,e}\left(\nabla\times{\bf B}\right)\times{\bf B}-\frac{T_e}{n}\nabla n^{\gamma_e}
\label{eq:efield2}
\end{equation}
where $n$ is the electron density,  ${\bf V}_i=\frac{1}{n}\sum^{\,N_{\rm sp}}_{j=1}{Z_j \int{f_j{\bf v}_j\mathrm{d}{\bf v}_j}}$ is the ion fluid velocity, and $Z_j=q_j/e$ is the relative charge of the ion species $j$ \citep{gargate+07}. Here we assumed that electrons have temperature $T_e$ and a polytropic equation of state with an  index $\gamma_e$ \citep{Lipatov02}.
Possible choices for $\gamma_e$ are the canonical value for monoatomic gases, $\gamma_e=5/3$, or an effective index $\gamma_{\rm eff}$ chosen to maintain  thermal equilibration between  protons and electrons.

For shocks  electrons and ions are  initialized with the same temperature upstream; 
then, if the polytropic index were set to the canonical value of 5/3, the electron pressure at the shock would increase only by a factor of $\Pi_e=r^{5/3}\simeq 4^{5/3}\sim 10$ for strong shocks, while the proton pressure jump is $\Pi_p\simeq 5/4 M_s^2\gtrsim \Pi_e$ \citep{LL5}.
Finally, $\gamma_{\rm eff}(M_s)$ can be  found by numerically solving the Rankine--Hugoniot jump conditions including the electron pressure with the additional condition $\Pi_e=\Pi_p$, i.e.,
\begin{equation}
[r(M_s)]^{\gamma_{\rm eff}}= \frac{5M_s^2-1}{4}, 
\end{equation}
where we posed the proton adiabatic index equal to 5/3.

The post-MHD terms  in  Eq.~\ref{eq:efield2}, such as the Hall term  ($\propto{\bf J}\times {\bf B}$) and the divergence of the electron pressure ($\propto \nabla n^{\gamma_e}$), allow magnetic reconnection to occur and capture most of the features of full kinetic approaches \citep{karimabadi+04}. 
An even more precise description of reconnection in the hybrid limit would also require the adoption of an anisotropic electron pressure tensor \citep[e.g.][]{le+09}, which goes beyond the scope of this paper.

For details on how the above equations are discretized on the staggered grid and on the algorithms used to solve them, we refer the reader to \S3 of the original dHybrid method paper \citep{gargate+07}. 
Since its construction, dHybrid has been used to study many aspects of non-relativistic shocks and non-thermal particle acceleration \citep{gs12,filam,DSA,MFA,diffusion}. 
While the algorithms in dHybrid do not preserve the solenoidality constraint on $\bf B$ to machine precision ---unlike most modern MHD codes and some hybrid-kinetic codes \citep[e.g.,][]{kunz+14} that use constrained transport on a staggered mesh--- we have never observed the resulting truncation error in $\nabla\cdot{\bf B}$ (which is typically smaller that the Poisson noise introduced by finite $N_{\rm ppc}$) to greatly influence the outcome of strongly non-linear problems such as the shocks we have studied here. 
This holds true across a wide range of resolutions, dimensionalities, and box sizes.

\begin{acknowledgments}
The authors would like to thank M. Kunz, P. Blasi, and E. Amato for many insightful discussions, and the anonymous referees for their comments, which led to an improved version of the manuscript. 
This research was supported by NASA (grant NNX17AG30G to DC), NSF (grant AST-1714658 to DC and AST-1517638 to AS), and Simons Foundation (grant 267233 to AS). 
Simulations were performed on computational resources provided by the Princeton's TIGRESS High-Performance Computing Center, the University of Chicago Research Computing Center, the NASA High-End Computing Program through the NASA Advanced Supercomputing Division at Ames Research Center, and XSEDE TACC (TG-AST100035 and TG-AST180008).
\end{acknowledgments}

\bibliographystyle{jpp}
\bibliography{Total_JR}

\end{document}